\documentclass[remotesensing,article,accept,pdftex,moreauthors]{Definitions/mdpi} 
\usepackage{makecell}
\usepackage{booktabs}
\usepackage{amsmath}    
\firstpage{1} 
\pubvolume{1}
\issuenum{1}
\articlenumber{0}
\pubyear{2026}
\copyrightyear{2026}
\externaleditor{Firstname Lastname} 
\datereceived{ } 
\daterevised{ } 
\dateaccepted{ } 
\datepublished{ } 

\Title{\highlighting{SIMBA:} 
 A Bidirectional Retrieval–Forward Simulation Framework for Modeling FY-4A/GIIRS Hyperspectral Infrared Radiances Toward NWP Applications}

\Author{\hl{Jingdong Shen} 
 $^{1}$,Fu Wang $^{2,*}$\orcidA{0000-0002-7540-3731}, Qifeng Lu$^{2}$, Hao Huang $^{1}$, Chunqiang Wu$^{2}$, Chi Yang $^{1}$ and Xiaofang Liu $^{1,*}$ }

\AuthorNames{Jingdong Shen, Fu Wang, Qifeng Lu, Hao Huang, Chunqiang Wu, Xiaofang Liu and Chi Yang}

\address{%
$^{1}$ \quad School of Computer Science and Engineering, Sichuan University of Science \& Engineering, Yibin 644000,  \hl{China}
;
\hl{324085406126@stu.suse.edu.cn} 
 (J.S.);
{324085406108@stu.suse.edu.cn} (H.H.);
{323085406228@stu.suse.edu.cn} (C.Y.)
\\
$^{2}$ State Key Laboratory of Severe Weather Meteorological Science and Technology, CMA Earth System Modeling and Prediction Centre and Key Laboratory of Earth System Modeling and Prediction, China Meteorological Administration, Beijing, 100081;
\href{mailto:wangfu@cma.cn}{wangfu@cma.cn} (F.W.)
\href{mailto:luqf@cma.gov.cn}{luqf@cma.gov.cn} (Q.L.)
\href{mailto:wucq@cma.gov.cn}{wucq@cma.gov.cn} (C.W.)

}

\corres{\hl{Correspondence:} 
 {wangfu@cma.cn} (F.W.); {lxf1969@suse.edu.cn} (X.L.)}

\abstract{%
Hyperspectral infrared observations are an important data source for numerical weather prediction (NWP) because they provide rich information on the vertical structure of atmospheric temperature and humidity. However, most existing deep learning methods mainly focus on one-way retrieval from radiances to atmospheric profiles, while the reverse radiance simulation process and the consistency between atmospheric state space and radiance observation space are insufficiently considered. In this study, we propose SIMBA, a unified bidirectional retrieval--forward simulation framework for FY-4A/GIIRS hyperspectral infrared radiance modeling toward NWP applications. The framework jointly performs atmospheric profile retrieval and radiance reconstruction, introduces a cycle-consistency constraint to strengthen the coupling between the two processes, and employs a bidirectional Mamba state-space module to capture long-range dependencies along pressure levels. Using collocated FY-4A/GIIRS observations and ERA5 reanalysis data, the proposed method is evaluated for temperature retrieval, specific humidity retrieval, long-wave radiance reconstruction, and medium wave radiance reconstruction. Experimental results show that SIMBA outperforms several representative deep learning baselines across both retrieval and reconstruction tasks, while ablation experiments confirm the contribution of the bidirectional design and cycle-consistency mechanism. These results demonstrate that the proposed framework is effective for joint atmospheric profile retrieval and hyperspectral infrared radiance modeling, and suggests potential for future Jacobian-related analysis and NWP-oriented extensions.}


\addhighlights{yes}
\renewcommand{\addhighlights}{%
\noindent\textbf{What are the main findings?}
 \begin{itemize}[labelsep=2.5mm,topsep=3pt]
 \item SIMBA, a cycle-consistent bidirectional framework inspired by the data assimilation of thw numerical weather prediction (NWP) system, is developed to jointly perform atmospheric profile retrieval and forward brightness temperature (BT) simulation for hyperspectral infrared observations.
 \item SIMBA achieves the best overall performance among the compared baseline models in temperature and specific humidity retrieval, and BT simulation, while improving vertical dependency modeling, observation-space consistency, and channel-wise stability across pressure levels and spectral regions.
 \end{itemize}\vspace{3pt}
\textbf{What are the implications of the main findings?}
 \begin{itemize}[labelsep=2.5mm,topsep=3pt]
 \item  The results indicate that joint retrieval and forward simulation is an effective way to build a more physically consistent relationship between hyperspectral observations and atmospheric states, making the framework better aligned with the needs of data assimilation of NWP.
 \item SIMBA includes a differentiable profile-to-radiance reconstruction branch that enables observation-space consistency assessment of retrieved atmospheric profiles. This validate-by-reconstruction strategy may provide a methodological reference for other retrieval problems in which estimated geophysical states need to remain consistent with their corresponding observations.
 \end{itemize}
}
\usepackage{float} 
\begin{document}


\section{Introduction}
The quality of numerical weather prediction (NWP) ~\cite{brotzge2023challenges} strongly depends on the accuracy of the initial atmospheric state, making the effective use of satellite observations~\cite{lei2025overview} essential for improving forecast skill. Among~available observation types, hyperspectral infrared measurements~\cite{carminati2019assessment} provide abundant information on the vertical structure of atmospheric temperature and water vapor. Compared with conventional radiosonde and surface observations, they offer broader spatial coverage~\cite{zhao2022overview}, higher temporal continuity, and~richer vertically resolved thermodynamic information, and~are therefore particularly valuable~\cite{okamoto2020assessment} over regions with sparse conventional~observations.

The Geostationary Interferometric Infrared Sounder (GIIRS) onboard FY-4A~\cite{yin2021impact,yang2017introducing} is the world’s first geostationary hyperspectral infrared sounder. Unlike polar-orbiting hyperspectral infrared instruments~\cite{chahine2006airs}, GIIRS provides high-frequency regional observations over fixed domains~\cite{schmit2017closer}, creating new opportunities for continuous monitoring of atmospheric thermodynamic conditions. However, GIIRS does not directly observe atmospheric temperature and humidity profiles, but~rather top-of-atmosphere infrared radiances. Therefore, a~key task is to establish an effective and stable mapping between atmospheric state variables and observed radiances~\cite{rodgers2000inverse}.

In conventional practice, this mapping is established through radiative transfer models (RTMs)~\cite{yang2020advanced,lu2020monitoring}, which connect atmospheric state space and observation space. RTMs play a central role in radiance simulation, profile retrieval, and~hyperspectral infrared applications in NWP~\cite{eyre2022assimilation}. In~particular, variational data assimilation~\cite{rabier2000ecmwf}  requires accurate forward operators and reliable sensitivity information to describe the relationship between atmospheric states and radiance observations~\cite{de2022coupled}. However, under~high-dimensional channel settings and large-scale data conditions, conventional RTMs can be computationally expensive, which motivates the development of more efficient surrogate modeling strategies~\cite{krasnopolsky2013application,krishnan2012artificial}.

Recent advances in deep learning have provided a new pathway for learning the nonlinear relationship between hyperspectral infrared radiances and atmospheric profiles~\cite{reichstein2019deep,chen2023machine,kashinath2021physics}. Existing studies have employed convolutional neural networks (CNNs)~\cite{li2023retrieving}, long short-term memory networks (LSTMs)~\cite{ye2021simultaneous}, Transformers~\cite{kong2026hybrid}, and~multilayer perceptrons (MLPs)~\cite{zhao2018improved} to estimate temperature or humidity profiles directly from radiance observations. \textcolor{black}{In the broader hyperspectral remote-sensing literature, generative adversarial networks (GANs) have also been used for spectral--spatial representation learning, data augmentation, and~hyperspectral image classification~\cite{ranjan2026dive}.} However, most of these studies mainly focus on one-way retrieval~\cite{malmgren2019statistical}, namely, learning the mapping from radiances to atmospheric profiles, while the forward simulation process from atmospheric states to radiances is rarely modeled explicitly within the same framework~\cite{lahoz2014data}. As~a result, model optimization is dominated by state-space errors, whereas consistency in observation space~\cite{gettelman2022future} is insufficiently constrained. \textcolor{black}{This limitation weakens the physical coherence of the learned mapping and reduces its ability to maintain consistency between atmospheric state space and radiance observation space.}~\cite{wang2024physics,bocquet2023surrogate}.

Another challenge arises from the intrinsic vertical structure of atmospheric thermodynamic profiles~\cite{kotthaus2022atmospheric}. Temperature and humidity profiles are organized along pressure levels and exhibit pronounced cross-level coupling and long-range dependencies~\cite{xue2022one}. Thus, hyperspectral infrared radiance--profile modeling is not only a nonlinear regression problem~\cite{koner2016deterministic}, but~also a vertical sequence modeling problem~\cite{xiao2024region}. Traditional recurrent neural networks may suffer from gradient attenuation when handling long sequences, whereas Transformer-based models often incur high computational cost due to global attention operations. Recently developed state-space models, especially the Mamba architecture, provide strong long-sequence modeling capability with linear computational complexity~\cite{gu2021efficiently}, making them well suited for representing multilayer vertical dependencies~\cite{chen2025mamba}.

To address these issues, we propose SIMBA, a~bidirectional retrieval--forward simulation framework for FY-4A/GIIRS hyperspectral infrared radiance modeling toward NWP applications~\cite{feng2022improving}. Instead of focusing only on atmospheric profile retrieval, we jointly model the radiance-to-profile retrieval process and the profile-to-radiance forward simulation process within a unified architecture~\cite{ye2025land}. Specifically, we design a retrieval branch to estimate atmospheric thermodynamic profiles from hyperspectral infrared radiances, and~a forward branch~\cite{gonzalez2024emulation} to reconstruct radiances from retrieved or reference profiles. We further couple the two branches through a cycle-consistency constraint~\cite{zhu2017unpaired}, so that the framework can be optimized in both state space and observation space simultaneously~\cite{cheng2021graph}. \textcolor{black}{In this way, we strengthen the linkage between atmospheric states and radiative observations and construct a data-driven differentiable forward-modeling framework that improves radiance–profile consistency under the evaluated conditions~\cite{brence2023surrogate,tahseen2024enhancing}.}
For this purpose, a~bidirectional Mamba state‑space module is incorporated into the framework to effectively capture the long‑range vertical dependencies inherent in atmospheric thermodynamic profiles, enabling more accurate modeling of the structured vertical coupling in temperature and~humidity.

The main contributions of this study are as follows:
(1) We propose a unified bidirectional retrieval--forward simulation framework for FY-4A/GIIRS hyperspectral infrared radiance modeling, which jointly links atmospheric state space and radiance observation space.
(2) We introduce a cycle-consistency constraint to couple the retrieval and forward branches, thereby improving observation-space consistency and the stability of bidirectional radiance--profile mapping.
(3) We incorporate a bidirectional Mamba state-space module to model long-range vertical dependencies of atmospheric thermodynamic profiles, and~demonstrate its effectiveness in both profile retrieval and radiance reconstruction tasks.
(4) \textcolor{black}{We provide a data-driven differentiable forward-modeling basis for future extensions toward sensitivity analysis and assimilation-oriented applications, while recognizing that rigorous RTM-based validation remains necessary.}

The remainder of this paper is organized as follows. Section~\ref{sec2} describes the data, sample construction strategy, and~methodology. Section~\ref{sec3} presents the experimental results and discussion. Section~\ref{sec4} concludes the paper and outlines future~work.

\section{Data and~Methodology}
\label{sec2}
\unskip

\subsection{Satellite and Reanalysis~Data}

This study uses observations from the Geostationary Interferometric Infrared Sounder (GIIRS) and AGRI onboard China’s FY-4A satellite~\cite{yang2017introducing}. GIIRS measures infrared radiances in two spectral bands: a long-wave (LW) band (700--1130\,cm\(^{-1}\), 689 channels) and a medium-wave (MW) band (1650--2250\,cm\(^{-1}\), 961 channels), both with a spectral resolution of 0.625\,cm\(^{-1}\). The~instrument operates with a \(32\times4\) field-of-view array scanning mode, providing a spatial resolution of approximately 16 km at the sub-satellite point and a \textcolor{black}{temporal resolution of 67 min for the China area and 35 min for the mesoscale area}. The~main technical specifications of GIIRS are listed in Table~\ref{tab1}. In~this study, Level 1 radiance products from the LW and MW bands are used together with AGRI Level 2 cloud-mask products and auxiliary information, including latitude, longitude, and~observation time. All GIIRS and AGRI data are obtained from the National Satellite Meteorological Centre (NSMC) of~China.

\begin{table}[H]

\caption{\hl{FY-4A} 
 GIIRS~specifications.}
\begin{tabularx}{\textwidth}{ll}
\toprule
\textbf{Parameter} & \textbf{Performance} \\
\midrule
Spectral bandwidth & 
\makecell[l]{Long-wave: 700--1130 cm\(^{-1}\) \\
            Medium-wave: 1650--2250 cm\(^{-1}\)} \\
\midrule
Spectral channels & 
\makecell[l]{Long-wave: 689 \\
            Medium-wave: 961} \\
\midrule
Spectral resolution & 
\makecell[l]{Long-wave: 0.625 cm\(^{-1}\) \\
            Medium-wave: 0.625 cm\(^{-1}\)} \\
\midrule
Sensitivity & 
\makecell[l]{Long-wave: 0.5--1.1 mW m\(^{-2}\) sr\(^{-1}\) cm\(^{-2}\) \\
            Medium-wave: 0.1--0.14 mW m\(^{-2}\) sr\(^{-1}\) cm\(^{-2}\)} \\
\midrule
Operational model & 
\makecell[l]{China area: 5000 $\times$ 5000 km\(^{2}\) \\
            Mesoscale area: 2000 $\times$ 2000 km\(^{2}\)} \\
\midrule
Spatial resolution & 16 km \\
\midrule
Temporal resolution & 
\makecell[l]{China area: 67 min \\
            Mesoscale area: 35 min} \\
\midrule
Calibration accuracy & 
\makecell[l]{1.5 K (3$\sigma$) radiation \\
            10 ppm (3$\sigma$) spectrum} \\
\bottomrule
\end{tabularx}
\label{tab1}
\end{table}

ERA5 reanalysis data~\cite{hersbach2020era5} (ECMWF) are used as reference atmospheric states for model training and evaluation. Temperature and specific humidity profiles at 101 vertical levels, from~the near surface to roughly 0.005 hPa, are used. \textcolor{black}{ ERA5 has a horizontal resolution of roughly 31 km and a temporal resolution of 60 min}, enabling precise spatiotemporal collocation with GIIRS observations. \textcolor{black}{ERA5 is used as a supervisory reference rather than as in situ ground truth, because~GIIRS provides footprint-scale radiances while ERA5 represents grid-scale fields. This scale mismatch may introduce representativeness errors into the training labels and force the model to map detailed observational features to smoothed targets, potentially degrading the representation of fine-scale or extreme localized events. The ERA5-based evaluation should therefore be interpreted as a consistency assessment with reanalysis profiles, not as independent in situ validation.}

\subsection{Matching Methods and~Dataset}

GIIRS provides 1650 spectral channels with rich but highly correlated information. Direct use of all channels would introduce substantial redundancy and increase computational cost~\cite{chang2020channel}. \textcolor{black}{Therefore, the~input channels used in this study were selected with reference to the previous GIIRS channel-selection study by Zhang et al.~\cite{202680}, rather than being newly optimized in the present study.} Based on Jacobian characteristics and cumulative influence coefficients, channels with high sensitivity to temperature and humidity were retained. In~the LW band (700--1130~cm$^{-1}$), channels in the CO$_2$ absorption region (700--790~cm$^{-1}$) and H$_2$O absorption region (779--951~cm$^{-1}$) were selected; in the MW band (1650--2250~cm$^{-1}$), channels in the H$_2$O absorption region (1650--2000~cm$^{-1}$) were retained, along with several window channels for near-surface thermodynamic information. This yields 106 LW and 99~MW channels, used in all~experiments.

Based on the selected channels, a~collocated dataset is constructed from FY-4A/GIIRS radiance observations and ERA5 reanalysis profiles, with~GIIRS as the reference. ERA5 profiles within $\pm$30~min are matched to the nearest spatial grid point and interpolated to the 101 pressure levels of the satellite product. Only samples with valid LW and MW radiances and corresponding ERA5 profiles are retained. \textcolor{black}{Cloudy-sky samples are selected because they represent a more challenging radiative regime with stronger nonlinear interactions among temperature, humidity, clouds, and~surface effects.} After spatiotemporal matching and quality control, 158,649 cloudy-sky samples are obtained. Each sample contains the selected radiance channels, pressure-level coordinates, and~the corresponding ERA5 temperature and humidity~profiles.

\textcolor{black}{To examine cross-condition generalization, a~clear-sky dataset is constructed using the same procedures, yielding 112,151 samples. The~cloudy-sky-trained models are evaluated on this dataset without retraining; results are reported in Section~\ref{sec3.1}.}

\subsection{Proposed~Network}

\begin{figure}[H]

\includegraphics[width=1\textwidth]{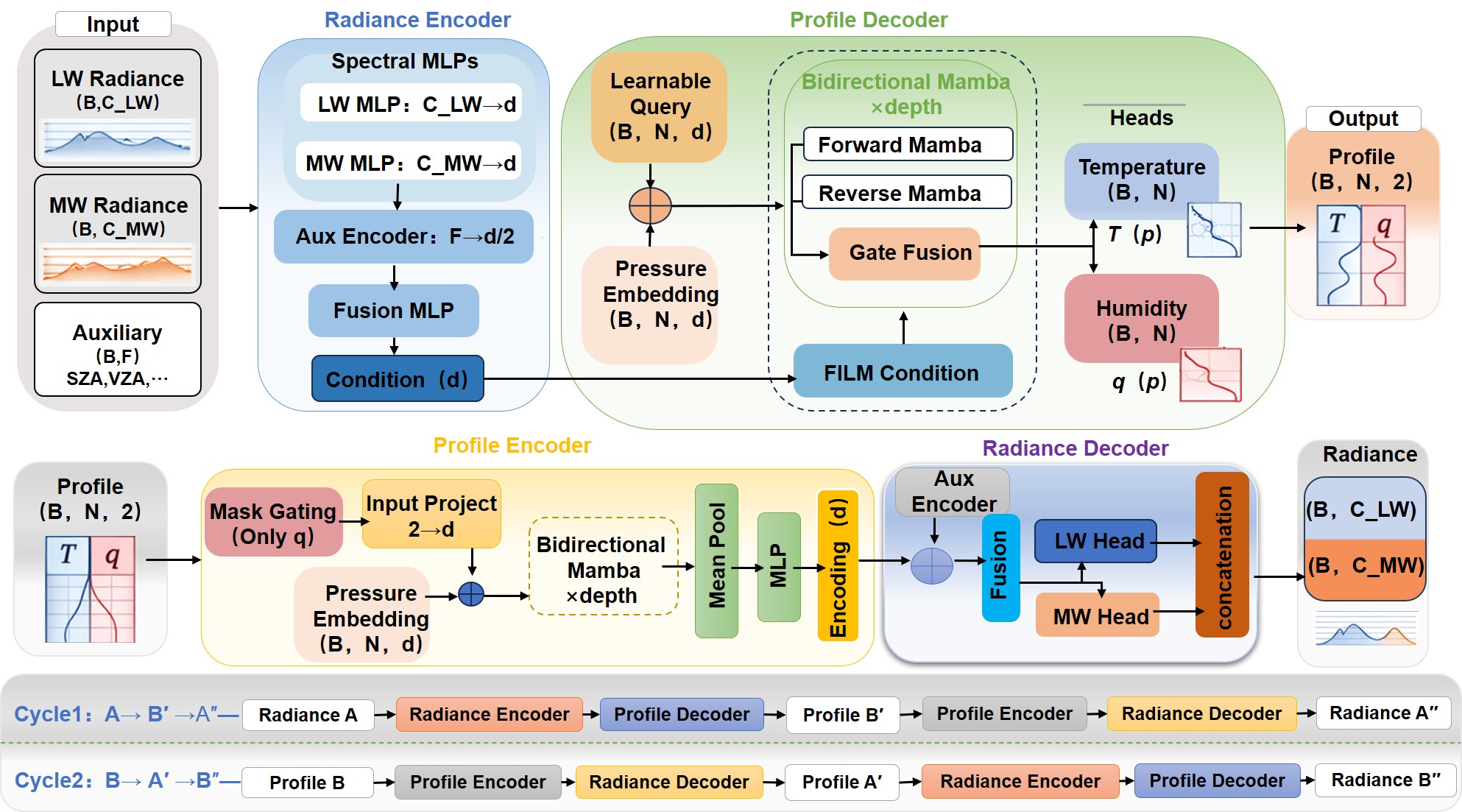}
\caption{\hl{Overall} 
 architecture of SIMBA. The~framework consists of a retrieval branch and a forward simulation branch for bidirectional radiance--profile modeling. A~bidirectional Mamba module is used for vertical sequence modeling, while FiLM conditioning injects radiative information into the profile generation process. The~two branches are coupled through a cycle-consistency~constraint.}
\label{fig1}
\end{figure}

In this study, we propose SIMBA, a~bidirectional retrieval--forward simulation framework for atmospheric profile retrieval and radiance reconstruction from FY-4A/GIIRS hyperspectral infrared observations. As~shown in Figure~\ref{fig1}, SIMBA consists of a retrieval branch and a forward simulation branch coupled through a cycle-consistency~mechanism.

SIMBA uses GIIRS LW and MW radiance observations together with auxiliary variables for radiance-to-profile retrieval, while atmospheric profiles are used in the forward branch for radiance reconstruction. ERA5 reanalysis profiles serve as supervisory reference states during~training.

\subsubsection{Retrieval and Forward Simulation~Branches}

In the retrieval branch, LW and MW radiances are first encoded by two spectral multilayer perceptron (MLP) encoders and then fused with auxiliary variables to form a unified conditional vector. The~fused features are passed to a profile decoder composed of stacked bidirectional Mamba blocks, which model the vertical dependencies of temperature and specific humidity along pressure levels. \textcolor{black}{Pressure-level positional encoding is introduced to preserve vertical structure. To~inject radiative information into the profile decoder, the~unified conditional vector is further used as the FiLM condition. Specifically, this vector summarizes the radiative and geometric information of each GIIRS sample and is passed through a fully connected layer to generate two sample-dependent modulation vectors, namely the scaling vector $\gamma(c)$ and the offset vector $\beta(c)$. These two vectors are then applied to the hidden states of the profile decoder, where $\gamma(c)$ adaptively adjusts the amplitude of the hidden features and $\beta(c)$ shifts their feature responses. In~this way, the~vertical profile generation process is dynamically conditioned on the radiative characteristics of each observation through a Feature-wise Linear Modulation (FiLM) mechanism~\cite{perez2018film}\hl{:} 
}
\begin{equation}
\mathrm{FiLM}(h,c)=\gamma(c)\odot h+\beta(c)
\end{equation}
where $h$ denotes the hidden state, $c$ denotes the conditional vector, and~$\gamma(\cdot)$ and $\beta(\cdot)$ are scaling and offset parameters generated by a fully connected~layer.

In the forward simulation branch, temperature and specific humidity profiles are encoded into latent atmospheric-state representations through a profile encoder with bidirectional Mamba blocks and pressure-level positional encoding. The~encoded features are then mapped to the LW and MW radiance spaces through a radiance decoder with band-specific output heads, enabling profile-to-radiance~reconstruction.

The two branches are coupled through a cycle-consistency constraint: profiles retrieved from radiances are required to reconstruct the original observations through the forward branch, while radiances reconstructed from reference profiles are required to recover the corresponding atmospheric profiles through the retrieval branch. This bidirectional closed-loop design improves the consistency and stability of radiance--profile~modeling.

\subsubsection{Loss~Function}

To jointly optimize the retrieval and forward simulation branches, we define a multi-objective loss function consisting of profile retrieval losses, radiance reconstruction losses, and~a cycle-consistency~loss.

The profile retrieval losses constrain the predicted temperature and specific humidity profiles using the corresponding ERA5 reference profiles:
\begin{equation}
\mathcal{L}_T = \frac{1}{N} \sum_{i=1}^N \| \hat{\mathbf{T}}_i - \mathbf{T}_i \|^2, \quad
\mathcal{L}_q = \frac{1}{N} \sum_{i=1}^N \| \hat{\mathbf{q}}_i - \mathbf{q}_i \|^2
\label{eq:profile_losses}
\end{equation}
where $\hat{\mathbf{T}}_i$ and $\hat{\mathbf{q}}_i$ denote the predicted temperature and specific humidity profiles, and~$\mathbf{T}_i$ and $\mathbf{q}_i$ denote the corresponding ERA5 reference~profiles.

The radiance reconstruction losses are defined for the LW and MW bands as
\begin{equation}
\mathcal{L}_{\mathrm{LW}} = \frac{1}{N} \sum_{i=1}^N \| \hat{\mathbf{R}}_i^{\mathrm{LW}} - \mathbf{R}_i^{\mathrm{LW}} \|^2, \quad
\mathcal{L}_{\mathrm{MW}} = \frac{1}{N} \sum_{i=1}^N \| \hat{\mathbf{R}}_i^{\mathrm{MW}} - \mathbf{R}_i^{\mathrm{MW}} \|^2
\label{eq:radiance_losses}
\end{equation}
where $\hat{\mathbf{R}}_i^{\mathrm{LW}}$ and $\hat{\mathbf{R}}_i^{\mathrm{MW}}$ denote the reconstructed LW and MW radiances, and~$\mathbf{R}_i^{\mathrm{LW}}$ and $\mathbf{R}_i^{\mathrm{MW}}$ denote the corresponding GIIRS~observations.

The cycle-consistency loss is defined as
\begin{equation}
\mathcal{L}_{\mathrm{cycle}} = \frac{1}{N} \sum_{i=1}^N \left( \| G(F(\mathbf{R}_i)) - \mathbf{R}_i \|^2 + \| F(G(\mathbf{P}_i)) - \mathbf{P}_i \|^2 \right),
\label{eq:cycle_loss}
\end{equation}
where $F$ denotes the retrieval mapping (radiance $\rightarrow$ profile), $G$ denotes the forward simulation mapping (profile $\rightarrow$ radiance), $\mathbf{R}_i$ denotes the input GIIRS radiance observation, and~$\mathbf{P}_i$ denotes the corresponding ERA5 reference~profile.

The total loss is formulated as
\begin{equation}
\mathcal{L}_{\mathrm{total}} =
\alpha_T \mathcal{L}_T +
\alpha_q \mathcal{L}_q +
\alpha_{\mathrm{LW}} \mathcal{L}_{\mathrm{LW}} +
\alpha_{\mathrm{MW}} \mathcal{L}_{\mathrm{MW}} +
\beta \mathcal{L}_{\mathrm{cycle}},
\label{eq:total_loss}
\end{equation}
where $\alpha_T$, $\alpha_q$, $\alpha_{\mathrm{LW}}$, $\alpha_{\mathrm{MW}}$, and~$\beta$ are hyperparameters controlling the relative contributions of the loss~terms. 

We implemented SIMBA and all baseline models using PyTorch 2.7.0.
The experimental dataset comprises FY-4A GIIRS radiation observations and ERA5 reanalysis profiles. Each sample contains preprocessed LW and MW radiation vectors, auxiliary information, pressure coordinates, and~corresponding reference temperature and specific humidity profiles. The~dataset was partitioned into training, validation, and~test sets at an 8:1:1 ratio, maintaining temporal sequence independence during splitting to objectively evaluate model generalization. Training employed the AdamW optimiser with an initial learning rate of 1 $\times$ 10$^{-3}$ and weight decay of 1 $\times$ 10$^{-4}$, combined with ReduceLROnPlateau learning rate scheduling to dynamically adjust the rate based on validation loss. Training was conducted on an NVIDIA RTX 4090 GPU with 24 GB of VRAM, employing a batch size of 128 and running for 100 epochs. All input data underwent standardization prior to~training.

\subsection{Evaluation~Metrics}

To quantitatively evaluate the performance of the proposed framework and baseline models, we used three standard regression metrics, namely root mean square error (RMSE), mean absolute error (MAE), and~the coefficient of determination ($R^2$). These metrics measure the magnitude of prediction errors and the overall goodness of fit. The~definitions are given as follows:
\begin{equation}
\mathrm{RMSE}=\sqrt{\frac{1}{n}\sum_{i=1}^{n}(\hat{y}_i-y_i)^2}
\end{equation}
\begin{equation}
\mathrm{MAE}=\frac{1}{n}\sum_{i=1}^{n}|\hat{y}_i-y_i|
\end{equation}
\begin{equation}
R^2 = 1-\frac{\sum_{i=1}^{n}(\hat{y}_i-y_i)^2}{\sum_{i=1}^{n}(y_i-\bar{y})^2}
\end{equation}
where $n$ is the number of samples, $y_i$ is the reference value, $\hat{y}_i$ is the model prediction, and~$\bar{y}$ is the sample mean of the reference values. In~the subsequent analysis, lower RMSE and MAE indicate better predictive accuracy, while an $R^2$ value closer to 1 indicates a better overall~fit.

\section{Results and~Discussion}
\label{sec3}
\unskip

\subsection{Overall~Evaluation}
\label{sec3.1}
\textcolor{black}{Bidirectional modeling has been applied in satellite data retrieval,  but~not yet in hyperspectral retrieval of temperature and humidity, where CNN~\cite{tan2024convolutional}, MLP~\cite{xu2024retrieval}, Transformer~\cite{xiao2024transformer},  and~LSTM~\cite{jiang2025deep}  remain widely used. We therefore augment these architectures with bidirectional processing, yielding Bidirectional CNN (BiCNN),  Bidirectional MLP (BiMLP),  Bidirectional Transformer (BiTransformer),  and~Bidirectional LSTM (BiLSTM) as comparison methods.}

Table~\ref{tab2} reports the mean absolute error (MAE), root mean square error (RMSE), and~coefficient of determination ($R^2$) for temperature and specific humidity profile retrieval. The~LW and MW radiance reconstruction errors are also included to evaluate the consistency between the retrieved atmospheric states and the observed radiances. \textcolor{black}{To examine cross-condition generalization, the~results are divided into cloudy-sky testing and clear-sky generalization, as~shown in Table~\ref{tab2}a and Table~\ref{tab2}b, respectively.} 

\begin{table}[H]

\renewcommand{\arraystretch}{1.25}
\setlength{\tabcolsep}{3pt}
\footnotesize
\caption{\hl{Comparative} 
 performance of SIMBA and the baseline models under cloudy-sky and clear-sky conditions. (\textbf{a}) Performance on the cloudy cases. (\textbf{b}) Performance on cases under clear-sky~conditions.}
\begin{tabularx}{\textwidth}{l c c c c c c c c c c c c}
\toprule
\multicolumn{13}{l}{\textbf{(a) Performance on the cloudy cases.}} \\
\midrule
\multirow{2}{*}{\vspace{-4pt}\hl{Model} 
} & \multicolumn{3}{c}{Temperature} & \multicolumn{3}{c}{Humidity} & \multicolumn{3}{c}{Long-wave} & \multicolumn{3}{c}{Medium-wave} \\
\cmidrule{2-13}
& MAE & RMSE & R$^2$ & MAE & RMSE & R$^2$ & MAE & RMSE & R$^2$ & MAE & RMSE & R$^2$ \\
\midrule
BiCNN         & 0.7399 & 1.0489 & \underline{0.9990} & 0.2601 & 0.5815 & 0.9891 & 4.0954 & 6.3157 & 0.8901 & \underline{0.2712} & \underline{0.3923} & \underline{0.8511} \\
BiMLP         & 0.8265 & 1.1911 & 0.9987 & 0.2753 & 0.6171 & 0.9877 & \underline{4.0434} & \underline{6.1790} & \underline{0.8948} & 0.2897 & 0.4099 & 0.8375 \\
BiTransformer & 0.7744 & 1.0890 & 0.9989 & 0.2650 & 0.5949 & 0.9886 & 4.2335 & 6.4741 & 0.8846 & 0.2800 & 0.4009 & 0.8445 \\
BiLSTM        & \underline{0.7338} & \underline{1.0346} & \underline{0.9990} & \underline{0.2578} & \underline{0.5772} & \underline{0.9892} & 4.1409 & 6.4350 & 0.8859 & 0.2774 & 0.4004 & 0.8450 \\
SIMBA         & \textbf{0.7131} & \textbf{1.0032} & \textbf{0.9991} & \textbf{0.2530} & \textbf{0.5684} & \textbf{0.9896} & \textbf{3.9674} & \textbf{6.1228} & \textbf{0.8967} & \textbf{0.2711} & \textbf{0.3887} & \textbf{0.8538} \\
\midrule
\multicolumn{13}{l}{\textbf{(b) Performance on cases under clear-sky conditions.}} \\
\midrule
\multirow{2}{*}{\vspace{-4pt}\hl{Model}} & \multicolumn{3}{c}{Temperature} & \multicolumn{3}{c}{Humidity} & \multicolumn{3}{c}{Long-wave} & \multicolumn{3}{c}{Medium-wave} \\
\cmidrule{2-13}
& MAE & RMSE & R$^2$ & MAE & RMSE & R$^2$ & MAE & RMSE & R$^2$ & MAE & RMSE & R$^2$ \\
\midrule
BiCNN         & 0.7955 & 1.3117 & 0.9984 & 0.4281 & 1.0801 & 0.9565 & 2.0381 & 2.9410 & 0.9851 & 0.2687 & 0.3637 & 0.9289 \\
BiMLP         & 0.9137 & 1.5644 & 0.9977 & 0.4522 & 1.1222 & 0.9530 & 1.9585 & 2.8969 & 0.9856 & 0.2885 & 0.3944 & 0.9164 \\
BiTransformer & 0.8152 & 1.3035 & 0.9984 & 0.4172 & 1.0473 & 0.9591 & 2.7880 & 4.0576 & 0.9717 & 0.3262 & 0.4465 & 0.8929 \\
BiLSTM        & \underline{0.7266} & \underline{1.1860} & \underline{0.9987} & \underline{0.4051} & \underline{1.0430} & \underline{0.9594} & \underline{1.7530} & \underline{2.5279} & \underline{0.9890} & \underline{0.2659} & \underline{0.3626} & \underline{0.9293} \\
SIMBA         & \textbf{0.6958} & \textbf{1.1006} & \textbf{0.9989} & \textbf{0.4016} & \textbf{1.0320} & \textbf{0.9603} & \textbf{1.6894} & \textbf{2.4016} & \textbf{0.9901} & \textbf{0.2614} & \textbf{0.3550} & \textbf{0.9323} \\
\bottomrule
\end{tabularx}
\label{tab2}
\end{table}

Table~\ref{tab2}a shows that SIMBA performs best under cloudy-sky conditions. For~temperature retrieval, it achieves an RMSE of 1.0032~K, an~MAE of 0.7131~K, and~an $R^2$ of 0.9991, reducing the RMSE by approximately 3.0\% relative to BiLSTM. For~specific humidity retrieval, SIMBA yields an RMSE of 0.5684, which is approximately 1.5\% lower than that of BiLSTM. Radiance reconstruction errors are also the lowest for both the LW and MW bands. \textcolor{black}{When evaluated on clear-sky samples (Table~\ref{tab2}b), SIMBA continues to outperform all baselines, with~RMSE reductions of 7.20\% for temperature, 1.05\% for specific humidity, 5.00\% for LW radiance, and~2.10\% for MW radiance relative to BiLSTM. 
This result indicates that cloudy-sky training does not restrict SIMBA to cloudy pixels; instead, the~bidirectional radiance--profile mapping generalizes to clear-sky conditions.}

Overall, SIMBA enhances retrieval and reconstruction under cloudy skies and generalizes to clear-sky conditions without retraining. These findings support the bidirectional framework and indicate that Mamba is well-suited for modeling vertical dependencies in atmospheric~profiles.

\textcolor{black}{We further evaluated the computational efficiency of SIMBA against all baseline models. All measurements were conducted on an NVIDIA RTX 4090 GPU (24 GB VRAM) with PyTorch 2.7.0, using identical preprocessing and inference settings. SIMBA has 1.13M trainable parameters—fewer than BiTransformer (1.76M) and comparable to BiLSTM (1.04M). Its inference time is 2.159 s for 13,785 test samples (0.157 ms per sample), and~its peak GPU memory usage is 146.50 MB. That is lower than BiLSTM (234.27 MB) but slightly higher than BiTransformer (130.60 MB). Inference speeds under cloudy-sky conditions are detailed in Table S3.}

\subsection{Evaluation of Atmospheric Profile~Retrieval } 

\subsubsection{Temperature~Profiles}

To further evaluate temperature profile retrieval performance, the~overall scatter distributions, layered scatter characteristics, and~vertical error profiles of different models are analyzed in this~section.

\begin{figure}[H]

\includegraphics[width=1\textwidth]{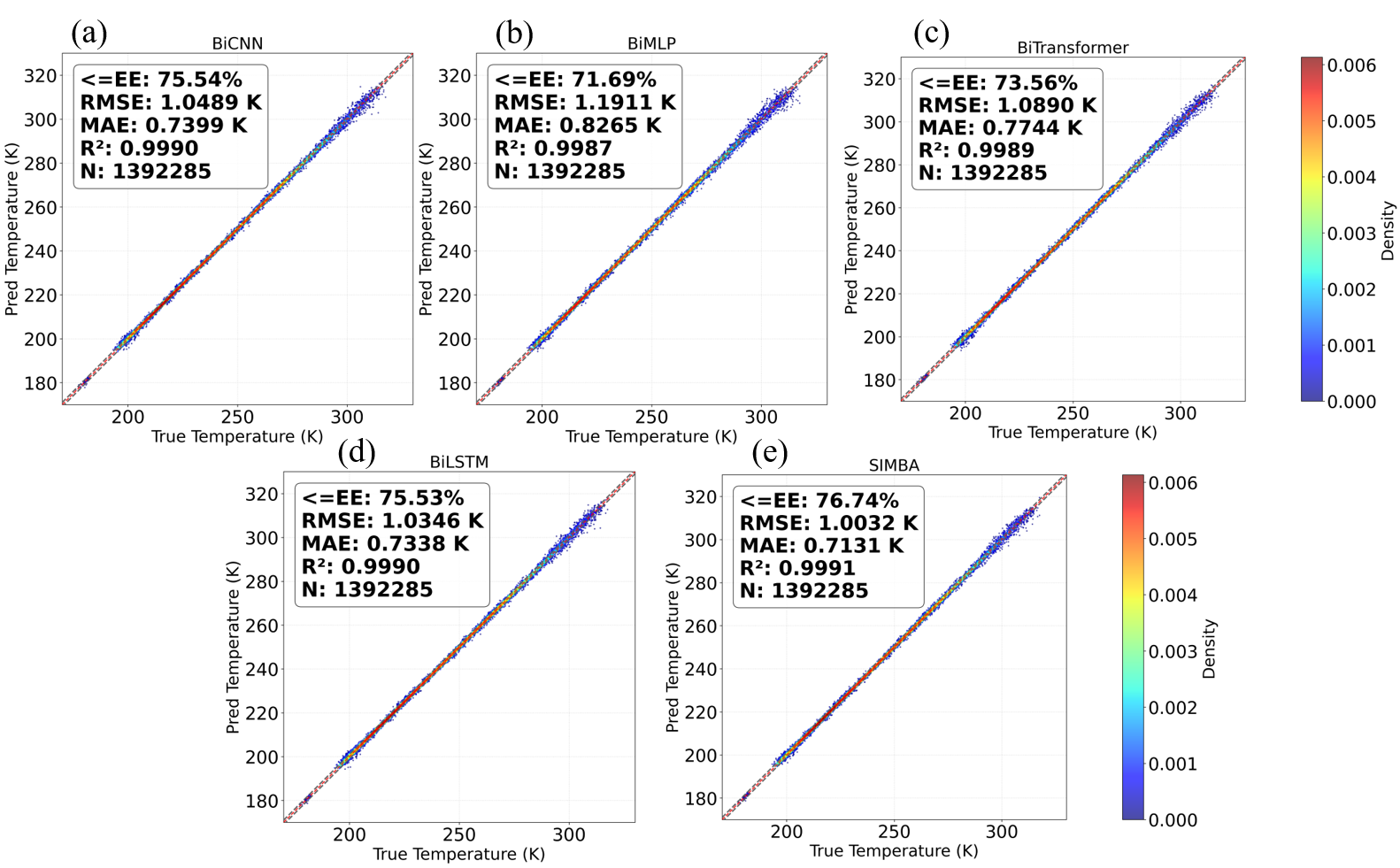}
\caption{\hl{Overall} 
 density scatter comparison of temperature retrievals on the test set for different models: (\textbf{a}) BiCNN; (\textbf{b}) BiMLP; (\textbf{c}) BiTransformer; (\textbf{d}) BiLSTM; and (\textbf{e}) SIMBA. The~red dashed line represents the 1:1 reference line, while the gray dashed lines represent the expected error (EE) envelope of $\pm 1$~K. The~color bar indicates the kernel density of the samples. The~metric $\leq$ EE indicates the proportion of temperature points within the EE envelope, i.e.,~$|T_{\mathrm{pred}} - T_{\mathrm{true}}| \leq 1$~K. $N$ denotes the total number of temperature points obtained by flattening all vertical levels of the test profiles.
}
\label{fig2}
\end{figure}

Figure~\ref{fig2} presents the overall density scatter distributions of temperature retrievals on the test set for different models. In~general, the~predictions of all models follow the 1:1 reference line, indicating that they can reproduce the overall temperature distribution reasonably well. \textcolor{black}{ To quantitatively characterize the scatter dispersion, an~expected error (EE) envelope of $\pm 1$ K was introduced around the 1:1 line, and~the percentage of temperature points within this envelope was calculated as the within-EE percentage. SIMBA achieves the highest within-EE value of 76.74\%, compared with 75.54\% for BiCNN, 71.69\% for BiMLP, 73.56\% for BiTransformer, and~75.53\% for BiLSTM. In~addition, SIMBA obtains the lowest RMSE of 1.0032 K and MAE of 0.7131 K, as~well as the highest $R^2$ of 0.9991. These quantitative results indicate that SIMBA has a more compact error distribution and better overall agreement with the reference temperature than the comparison models.}

\begin{figure}[H]

\includegraphics[width=1\textwidth]{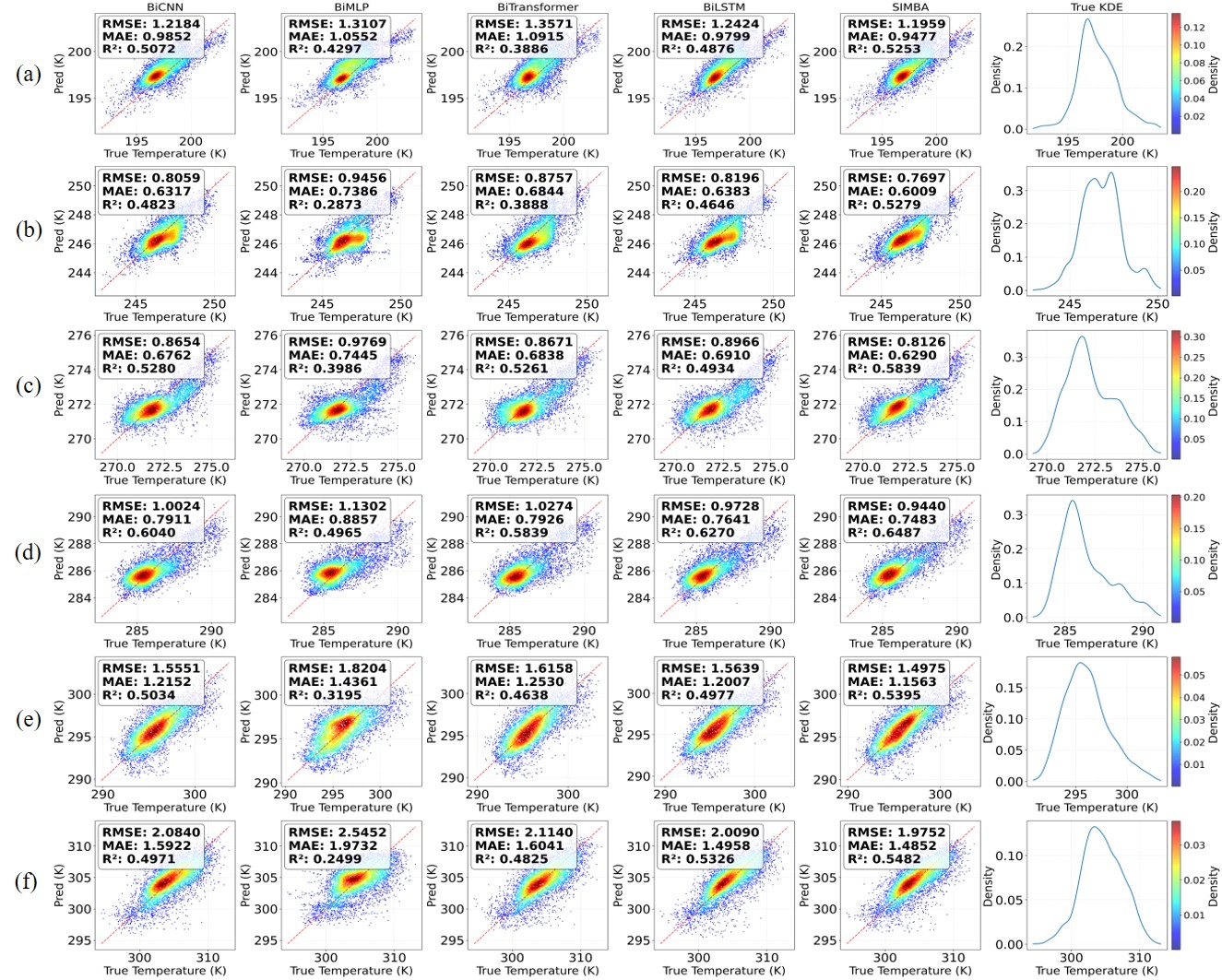}
\caption{Scatter plot comparison of temperature retrieval at six representative pressure levels: (\textbf{a})~100~hPa; (\textbf{b}) 300 hPa; (\textbf{c}) 500 hPa; (\textbf{d}) 700 hPa; (\textbf{e}) 850 hPa; (\textbf{f}) 1000 hPa. Columns correspond to BiCNN, BiMLP, BiTransformer, BiLSTM, and~SIMBA, while the rightmost column shows the kernel density estimate (True KDE) of ERA5 temperature at each pressure~level.}
\label{fig3}
\end{figure}

The layered scatter plots in Figure~\ref{fig3} further illustrate temperature retrieval performance at six representative pressure levels (100, 300, 500, 700, 850, and~1000 hPa). Overall, the~predicted values of all models generally follow the 1:1 reference line, but~SIMBA shows a more concentrated distribution and reduced scatter at most pressure levels. The~advantage is particularly evident in the near-surface layer (850--1000 hPa) and the upper-level region (100--300 hPa), where the retrieval task is more~challenging.

\begin{figure}[H]

\includegraphics[width=1\textwidth]{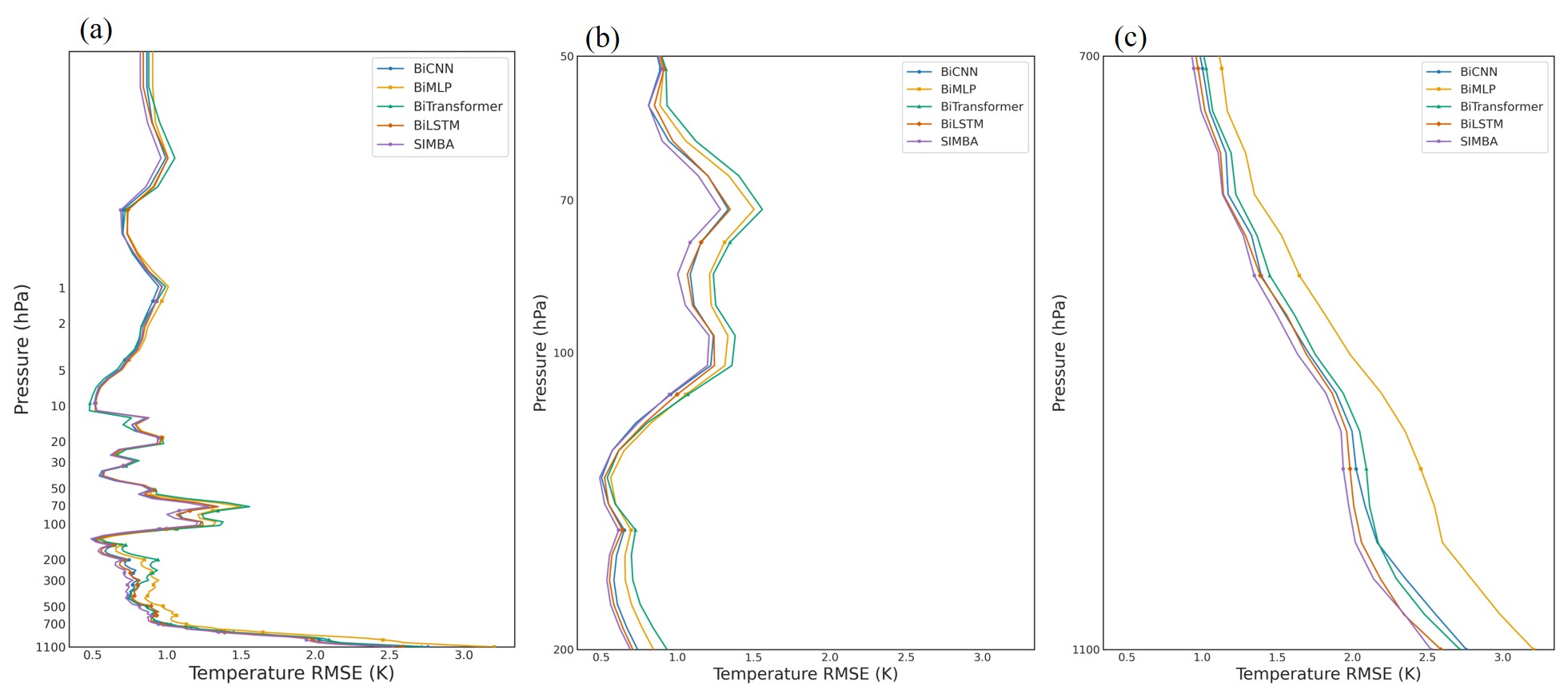}
\caption{Vertical distribution of RMSE for temperature retrieval across different models: (\textbf{a}) full pressure range; (\textbf{b}) enlarged view of 50--200 hPa; (\textbf{c}) enlarged view of 700--1100~hPa.}
\label{fig4}
\end{figure}

\begin{figure}[H]

\includegraphics[width=1\textwidth]{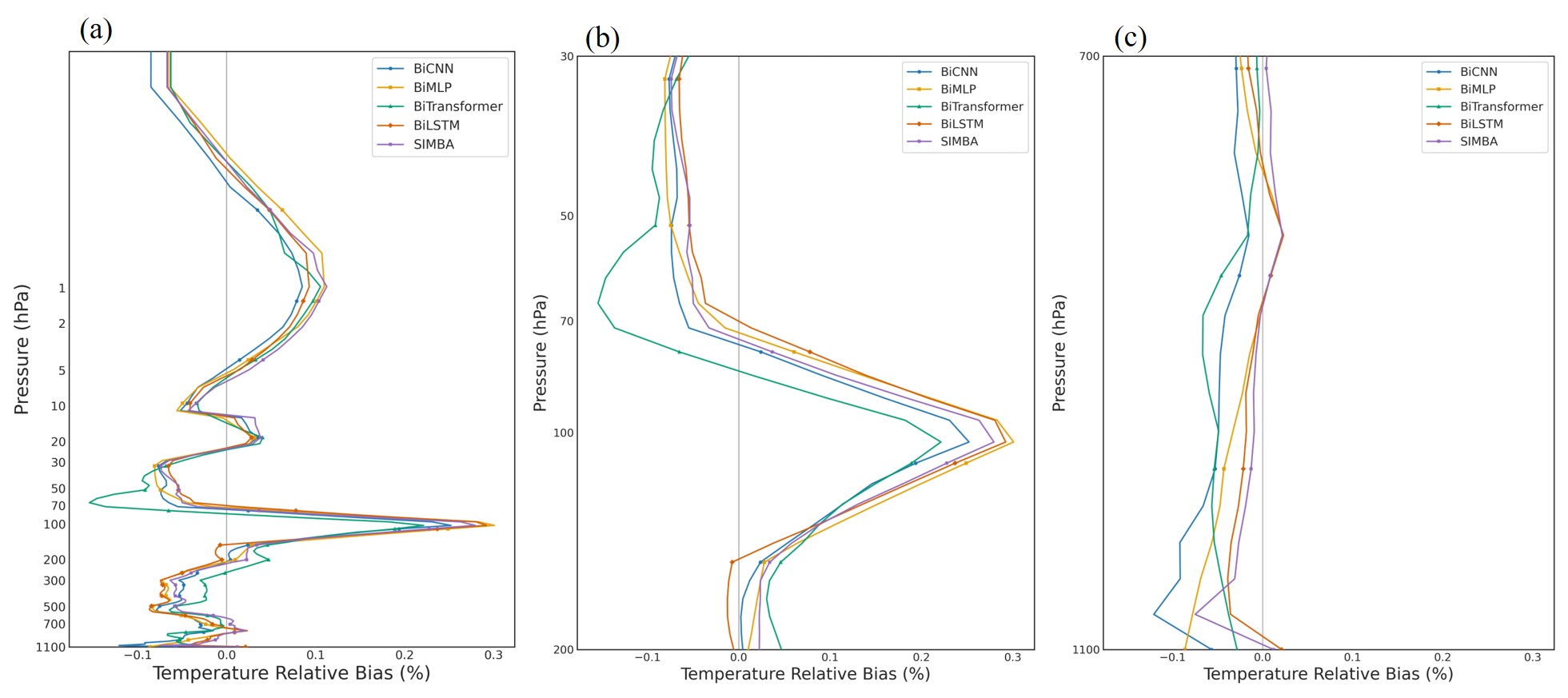}
\caption{Vertical distribution of relative bias for temperature retrieval across different models: (\textbf{a})~full pressure range; (\textbf{b}) enlarged view of 30--200 hPa; (\textbf{c}) enlarged view of 700--1100~hPa.}
\label{fig5}
\end{figure}

Figures~\ref{fig4} and~\ref{fig5} show the vertical distributions of RMSE and relative bias for temperature retrieval. Overall, SIMBA maintains comparatively low RMSE over most pressure levels, with~more noticeable improvements below 700 hPa and above 300 hPa. The~relative bias profiles further show that SIMBA exhibits smaller oscillation amplitudes throughout the vertical column, indicating a more stable error distribution across pressure levels. \textcolor{black}{In addition, we conducted an independent validation using 83 collocated radiosonde profiles. Between~200~hPa and 800~hPa, the~mean temperature bias is merely $-0.1637~\mathrm{K}$ with a standard deviation of $1.6833~\mathrm{K}$, while the mean specific-humidity bias is $0.1812~\mathrm{g~kg^{-1}}$ with a standard deviation of $1.3435~\mathrm{g~kg^{-1}}$, as~detailed in Figure~S2.}

Overall, the~temperature retrieval results demonstrate that SIMBA provides more accurate and vertically stable predictions than the comparison models. These results further support the effectiveness of the proposed bidirectional Mamba framework for modeling the vertical structure of atmospheric temperature~profiles.

\unskip

\subsubsection{Humidity~Profiles}

To further evaluate specific humidity retrieval performance, the~overall scatter distributions, layered scatter characteristics, and~vertical error profiles of different models are analyzed in this section. Compared with temperature retrieval, specific humidity retrieval is generally more challenging because of the stronger nonlinearity and vertical variability of water~vapor.

\begin{figure}[H]

\includegraphics[width=1\textwidth]{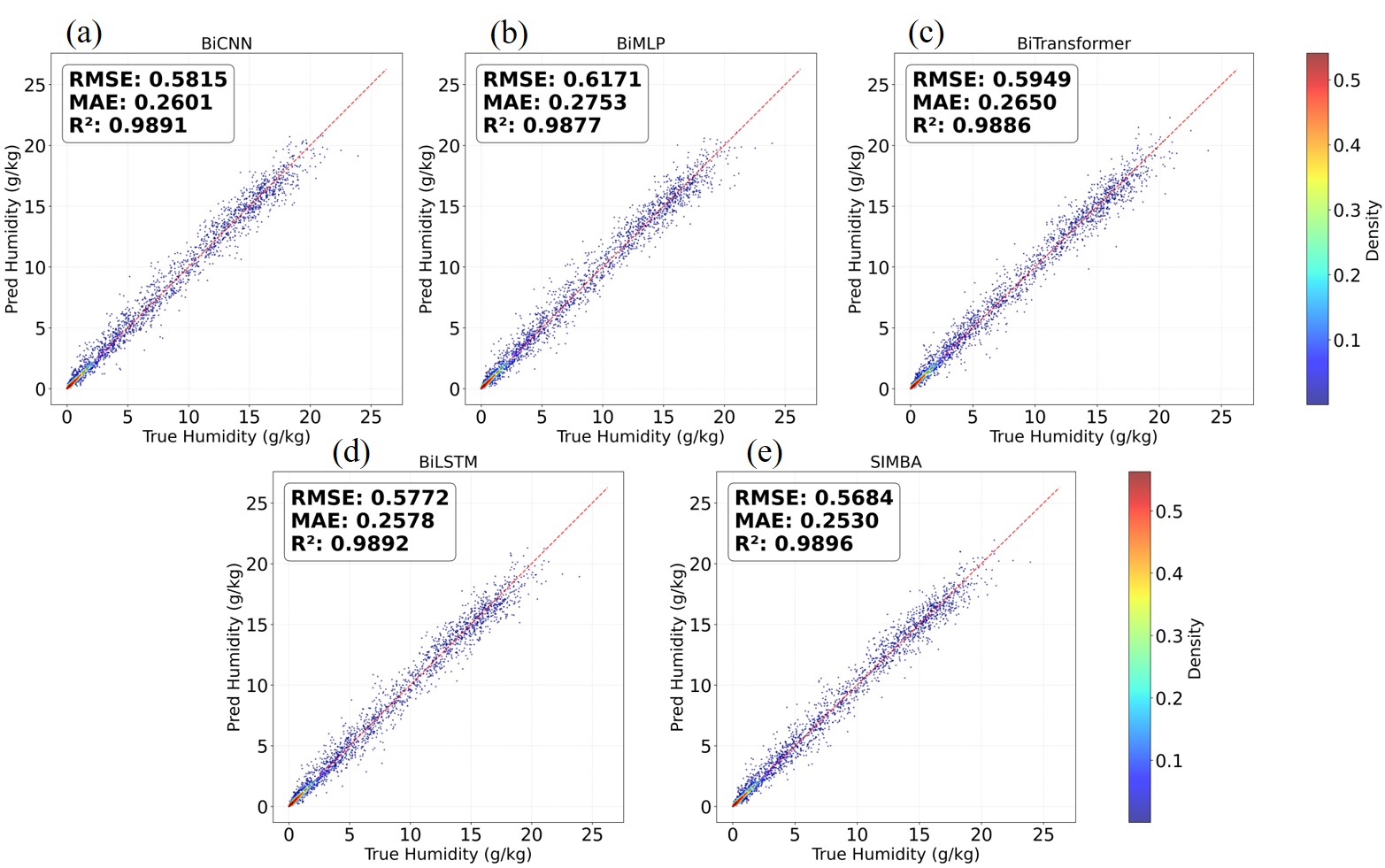}
\caption{Overall scatter plot comparison of specific humidity retrieval on the test set for different models: (\textbf{a}) BiCNN; (\textbf{b}) BiMLP; (\textbf{c}) BiTransformer; (\textbf{d}) BiLSTM; (\textbf{e}) SIMBA.}
\label{fig6}
\end{figure}

Figure~\ref{fig6} presents the overall scatter distributions of specific humidity retrieval on the test set for different models. In~general, the~predictions of all models follow the 1:1 reference line, indicating that they can reproduce the overall humidity distribution reasonably well. However, SIMBA exhibits a narrower scatter dispersion and achieves the lowest overall RMSE and MAE, indicating superior fitting performance for specific humidity retrieval. Compared with temperature retrieval, the~scatter becomes more dispersed in the high-humidity range (greater than 15 g/kg), reflecting the increased difficulty of humidity inversion. This visual pattern is consistent with the quantitative results in Table~\ref{tab2}, where SIMBA achieves the lowest specific humidity RMSE (0.5684) and MAE (0.2530), corresponding to an RMSE reduction of approximately 1.5\% relative to the best baseline, BiLSTM.

\begin{figure}[H]

\includegraphics[width=1\textwidth]{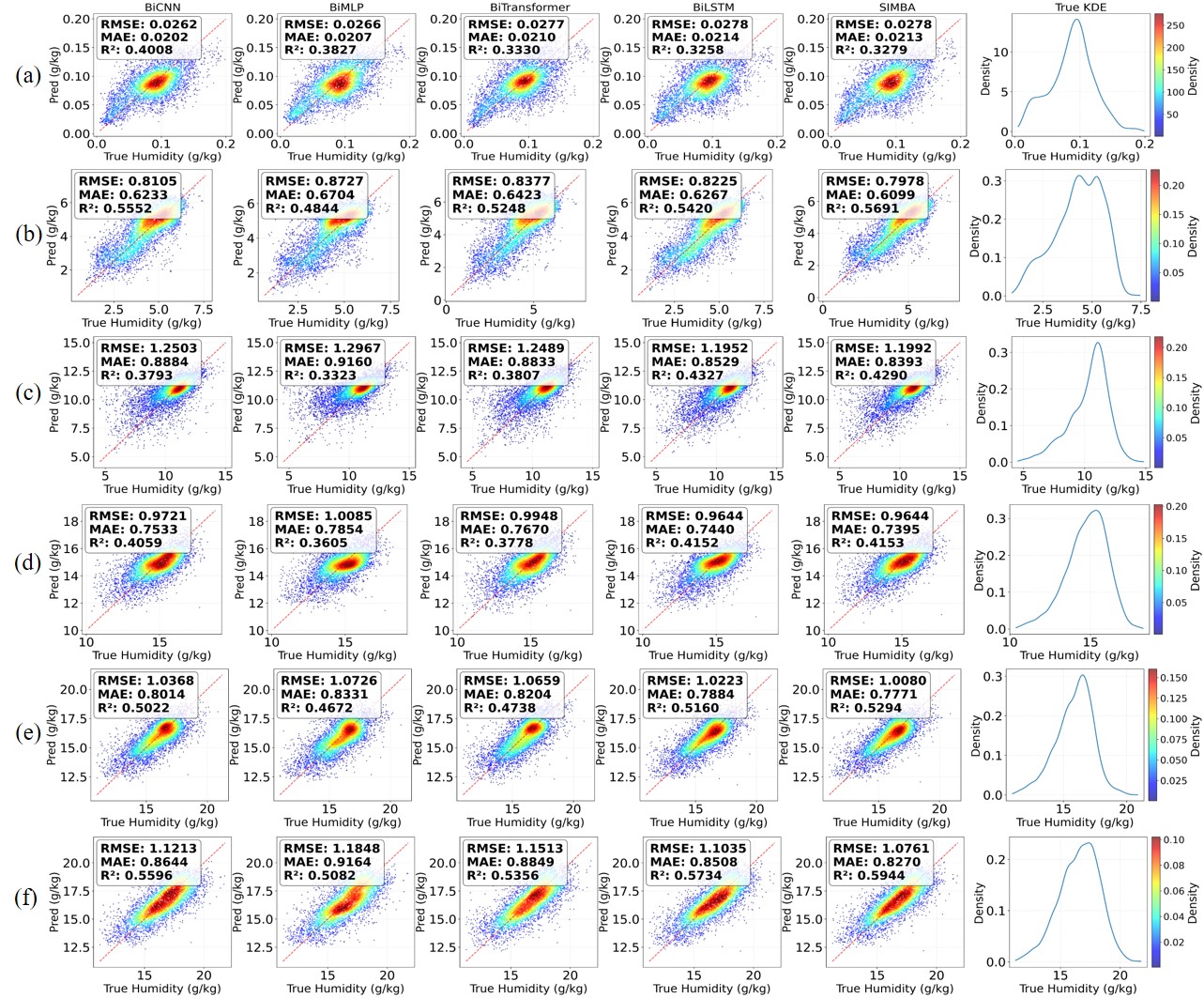}
\caption{Scatter plot comparison of specific humidity retrieval at six representative pressure levels: (\textbf{a}) 200 hPa; (\textbf{b}) 500 hPa; (\textbf{c}) 700 hPa; (\textbf{d}) 850 hPa; (\textbf{e}) 925 hPa; (\textbf{f}) 1000 hPa. Columns correspond to BiCNN, BiMLP, BiTransformer, BiLSTM, and~SIMBA, while the rightmost column shows the kernel density estimate (True KDE) of ERA5 specific humidity at each pressure~level.}
\label{fig7}
\end{figure}

The layered scatter plots in Figure~\ref{fig7} further demonstrate retrieval performance at six representative pressure levels (200, 500, 700, 850, 925, and~1000 hPa). Overall, SIMBA shows a more concentrated scatter distribution and lower overall error than most baseline models in the mid-to-lower troposphere, especially around 700 hPa and in the near-surface layers (850--1000 hPa), where humidity gradients are relatively strong and retrieval is more challenging. In~the upper layer (around 200 hPa), the~dispersion of all models increases, reflecting the greater uncertainty associated with the very low water vapor content at these altitudes. \textcolor{black}{ Specifically, at~(a) 200 hPa, SIMBA achieves an MAE of 0.0213 g/kg and an RMSE of 0.0278 g/kg; at (b) 500 hPa, the~MAE and RMSE increase to 0.6099 and 0.7978 g/kg, respectively. At~(e) 925 hPa and (f) 1000 hPa, SIMBA yields MAE values of 0.7771 and 0.8270 g/kg with RMSE values of 1.0080 and 1.0761 g/kg, consistently outperforming all baseline models in these high-humidity regimes.}

\begin{figure}[H]

\includegraphics[width=1\textwidth]{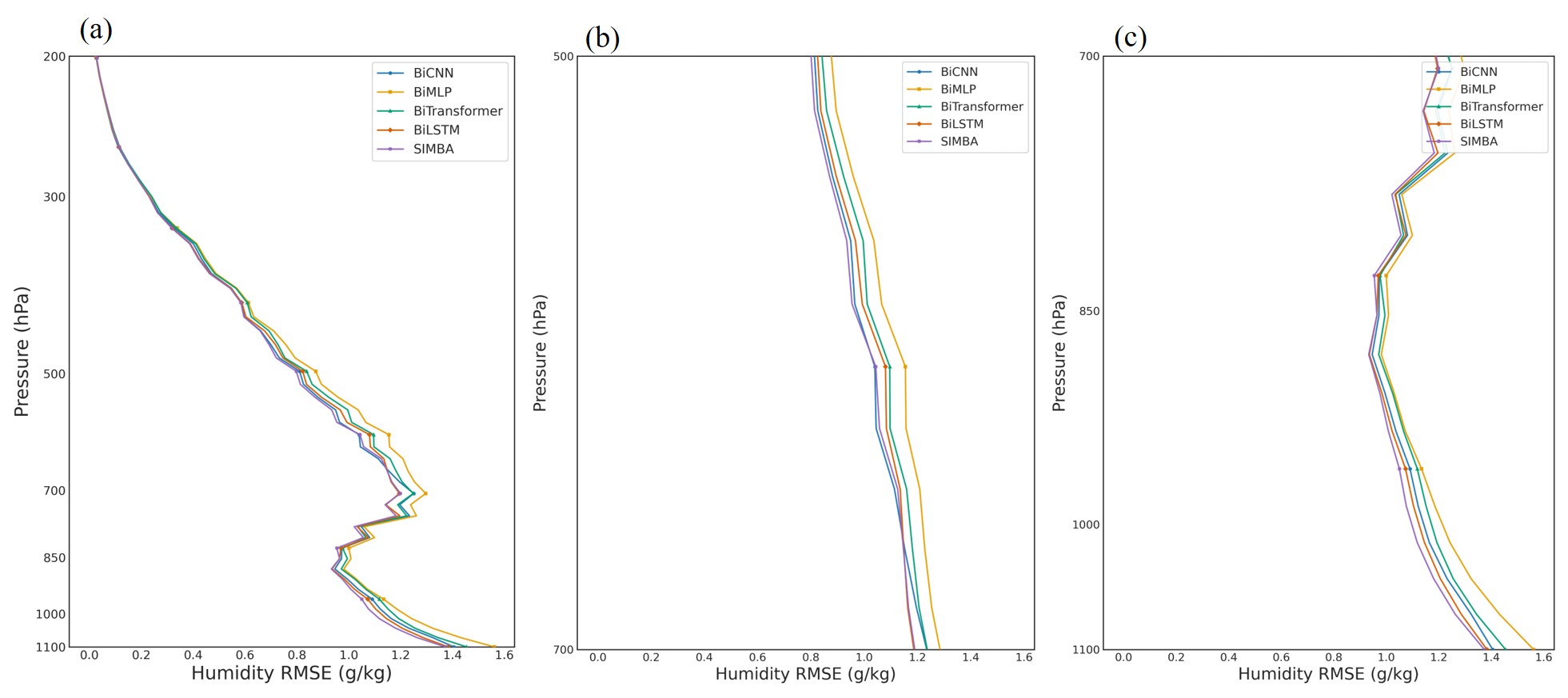}
\caption{Vertical distribution of RMSE for specific humidity retrieval across different models: (\textbf{a}) full pressure range; (\textbf{b}) enlarged view of 500--700 hPa; (\textbf{c}) enlarged view of 700--1100~hPa.}
\label{fig8}
\end{figure}

\begin{figure}[H]

\includegraphics[width=1\textwidth]{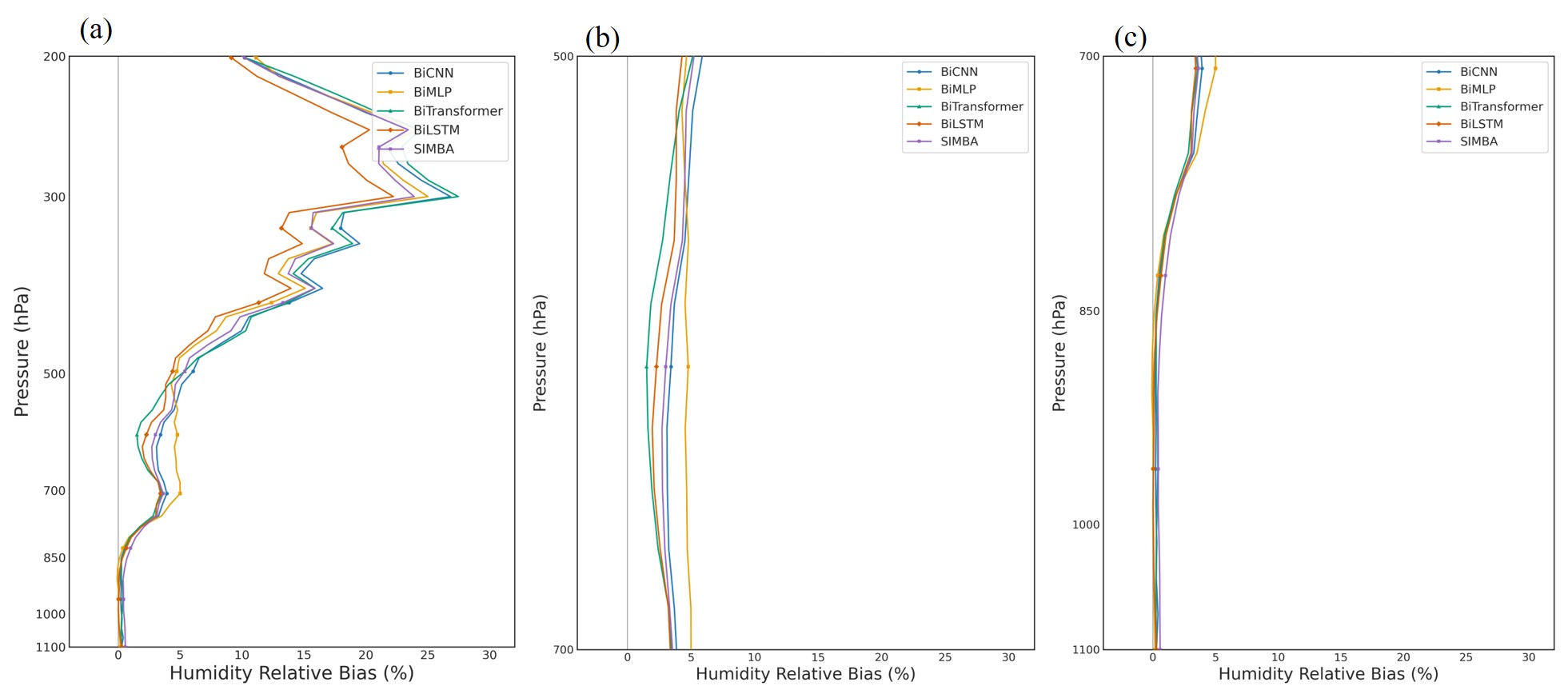}
\caption{Vertical distribution of relative bias for specific humidity retrieval across different models: (\textbf{a}) full pressure range; (\textbf{b}) enlarged view of 200--500 hPa; (\textbf{c}) enlarged view of 700--1100~hPa.}
\label{fig9}
\end{figure}

Figures~\ref{fig8} and~\ref{fig9} present the vertical distributions of RMSE and relative bias for specific humidity retrieval. Overall, SIMBA maintains comparatively low RMSE over most pressure levels, with~particularly notable improvements in the lower troposphere between 700 and 1000 hPa. The~relative bias profiles further show that SIMBA exhibits smaller oscillation amplitudes over the vertical column, indicating a more stable error distribution across pressure~levels.

Overall, the~humidity retrieval results demonstrate that SIMBA provides more accurate and vertically stable predictions than the comparison models, particularly in the moisture-rich lower troposphere. These results further support the effectiveness of the proposed bidirectional Mamba framework for modeling the vertically coupled structure of atmospheric moisture~profiles.

\vspace{-9pt}

\unskip

\vspace{-9pt}

\subsection{Evaluation for Forward~Models}
\unskip

\subsubsection{Radiation of Long-Wave~Bands}

From the perspective of NWP data assimilation, forward consistency in observation space is important for assessing whether the retrieved atmospheric states remain compatible with the observed radiance field. Therefore, the~LW radiance reconstruction results are analyzed in this~section.

\begin{figure}[H]

\includegraphics[width=1\textwidth]{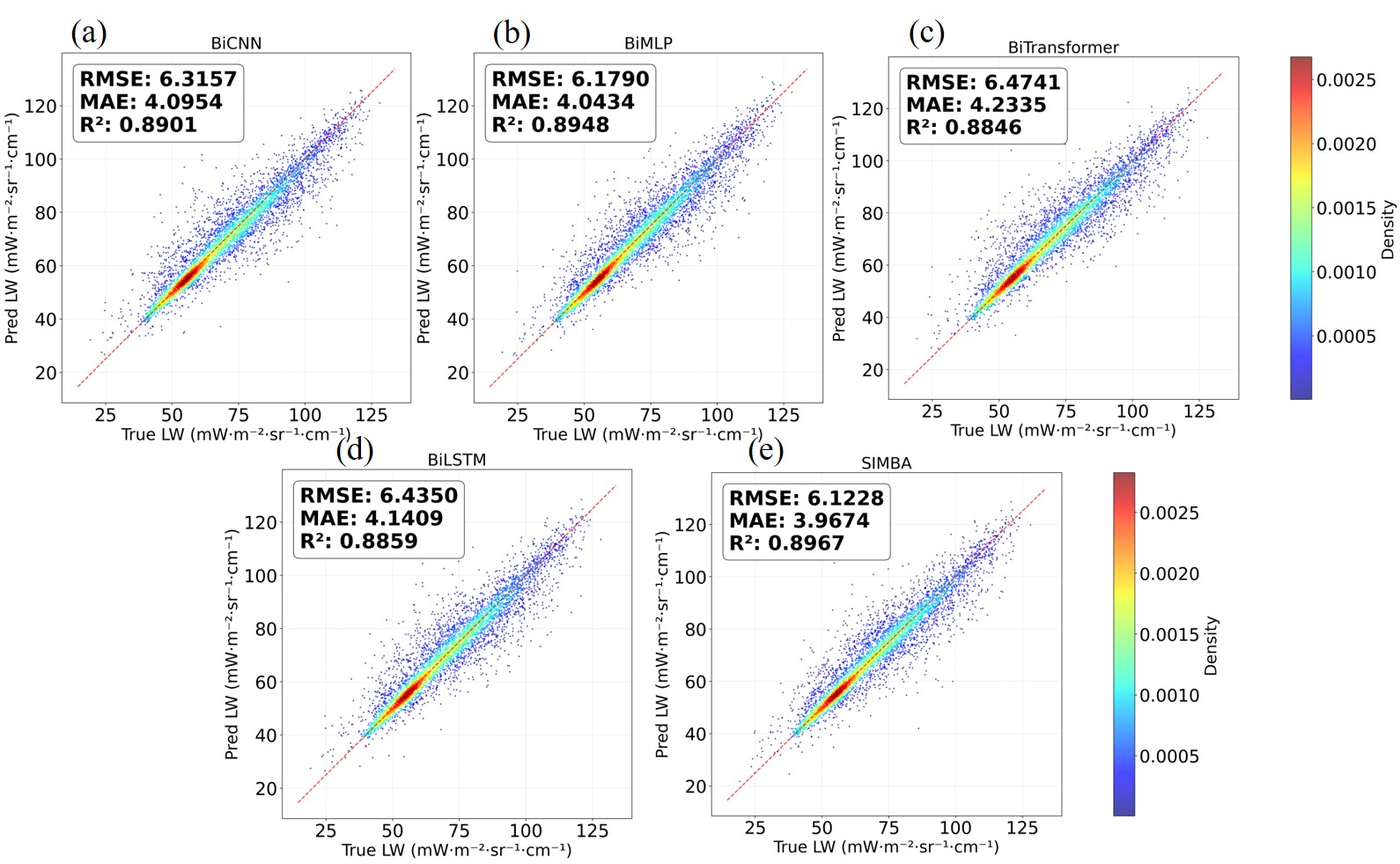}
\caption{\hl{Overall} 
 scatter plot comparison of long-wave radiance reconstruction on the test set for different models: (\textbf{a}) BiCNN; (\textbf{b}) BiMLP; (\textbf{c}) BiTransformer; (\textbf{d}) BiLSTM; (\textbf{e}) SIMBA.}
\label{fig10}
\end{figure}

Figure~\ref{fig10} presents the overall scatter distributions of reconstructed LW radiances on the test set for different models. In~general, the~reconstructed values of all models follow the 1:1 reference line, indicating that the retrieved atmospheric profiles can reproduce the main characteristics of the observed LW radiances. However, SIMBA exhibits a more concentrated scatter distribution and a narrower dispersion range than the comparison models, and~achieves the best overall performance in terms of RMSE, MAE, and~$R^2$. Although~the numerical improvements in LW reconstruction are smaller than those in profile retrieval, the~more compact scatter pattern indicates better consistency between the retrieved profiles and the observed~radiances.

\begin{figure}[H]

\includegraphics[width=1\textwidth]{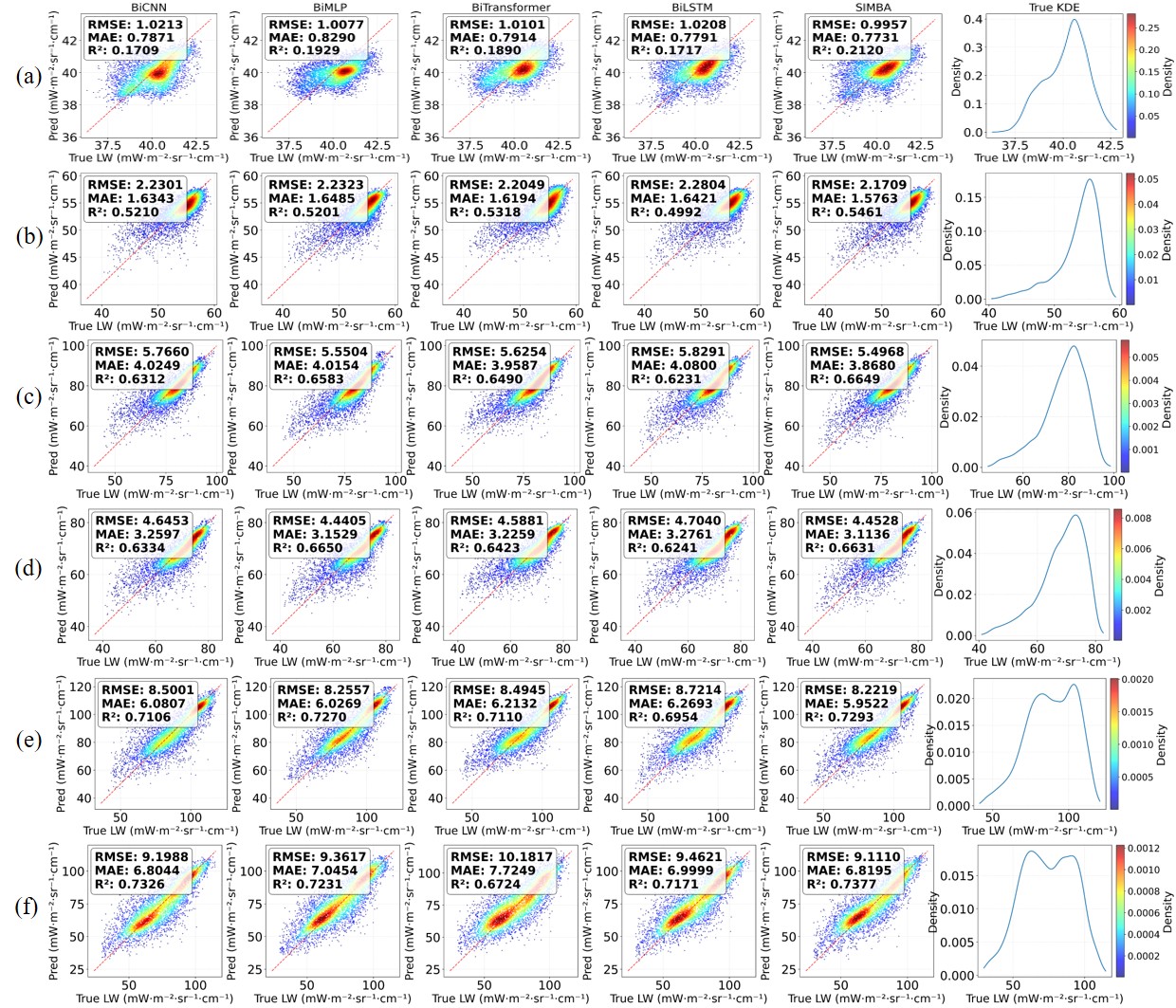}
\caption{\hl{Scatter} 
 plot comparison of long-wave radiance reconstruction for six representative channels: (\textbf{a}) CH1; (\textbf{b}) CH20; (\textbf{c}) CH35; (\textbf{d}) CH60; (\textbf{e}) CH80; (\textbf{f}) CH100. Columns correspond to BiCNN, BiMLP, BiTransformer, BiLSTM, and~SIMBA, while the rightmost column shows the kernel density estimate (True KDE) of the observed radiance at each~channel.}
\label{fig11}
\end{figure}

To further examine reconstruction differences across distinct spectral positions, six representative LW channels (CH1, CH20, CH35, CH60, CH80, and~CH100) are selected for detailed analysis in Figure~\ref{fig11}. Channels in the window region (e.g., CH1 and CH20) are more sensitive to near-surface thermodynamic structure, and~all models show relatively good linear relationships in these channels. By~contrast, channels within or near the CO$_2$ absorption region (e.g., CH60, CH80, and~CH100) are more sensitive to the middle and upper troposphere, and~the scatter distributions become more dispersed across all models. Nevertheless, SIMBA maintains comparatively lower dispersion and smaller systematic deviation in these absorption-sensitive~channels.

\begin{figure}[H]

\includegraphics[width=1\textwidth]{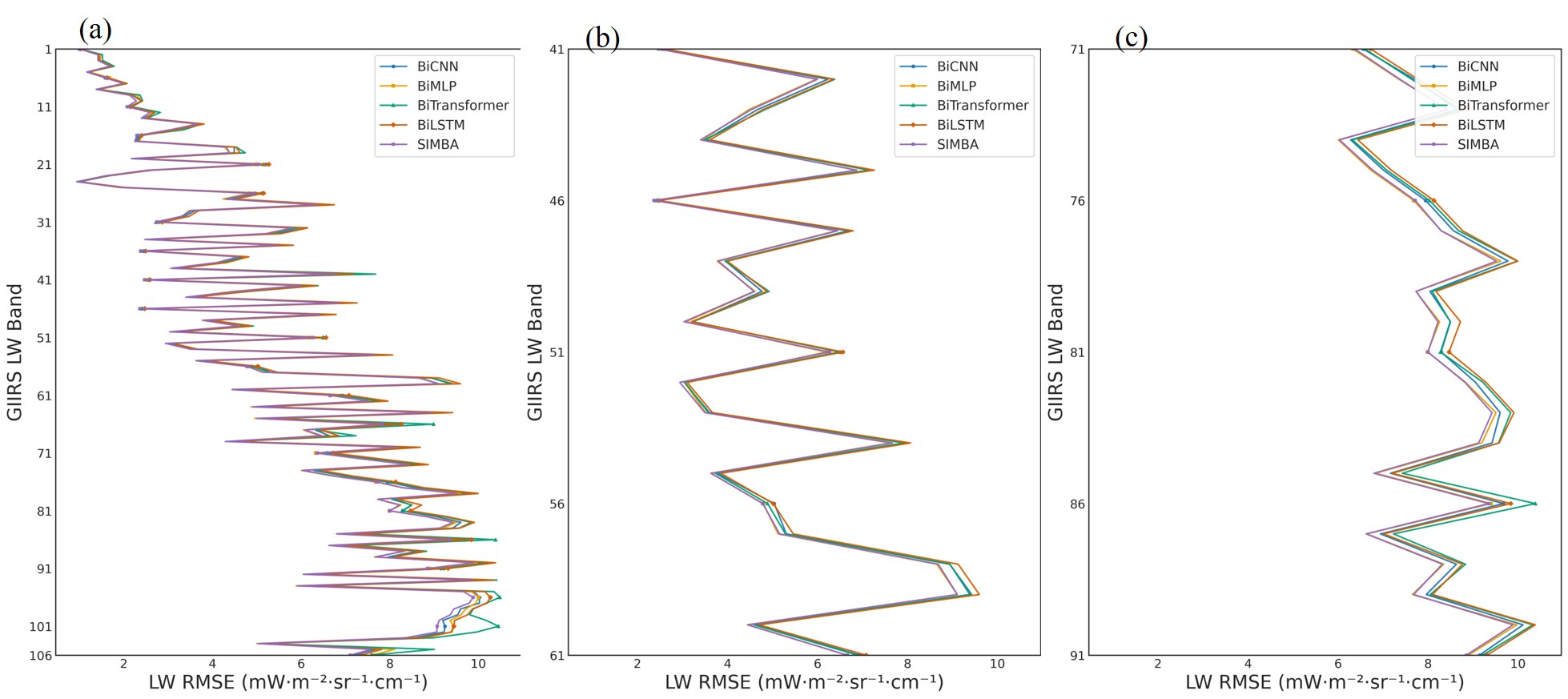}
\caption{Distribution of RMSE across 106 long-wave channels for different models: (\textbf{a}) full channel range; (\textbf{b}) enlarged view of CH41--CH61; (\textbf{c}) enlarged view of~CH71--CH91.}
\label{fig12}
\end{figure}

\begin{figure}[H]

\includegraphics[width=1\textwidth]{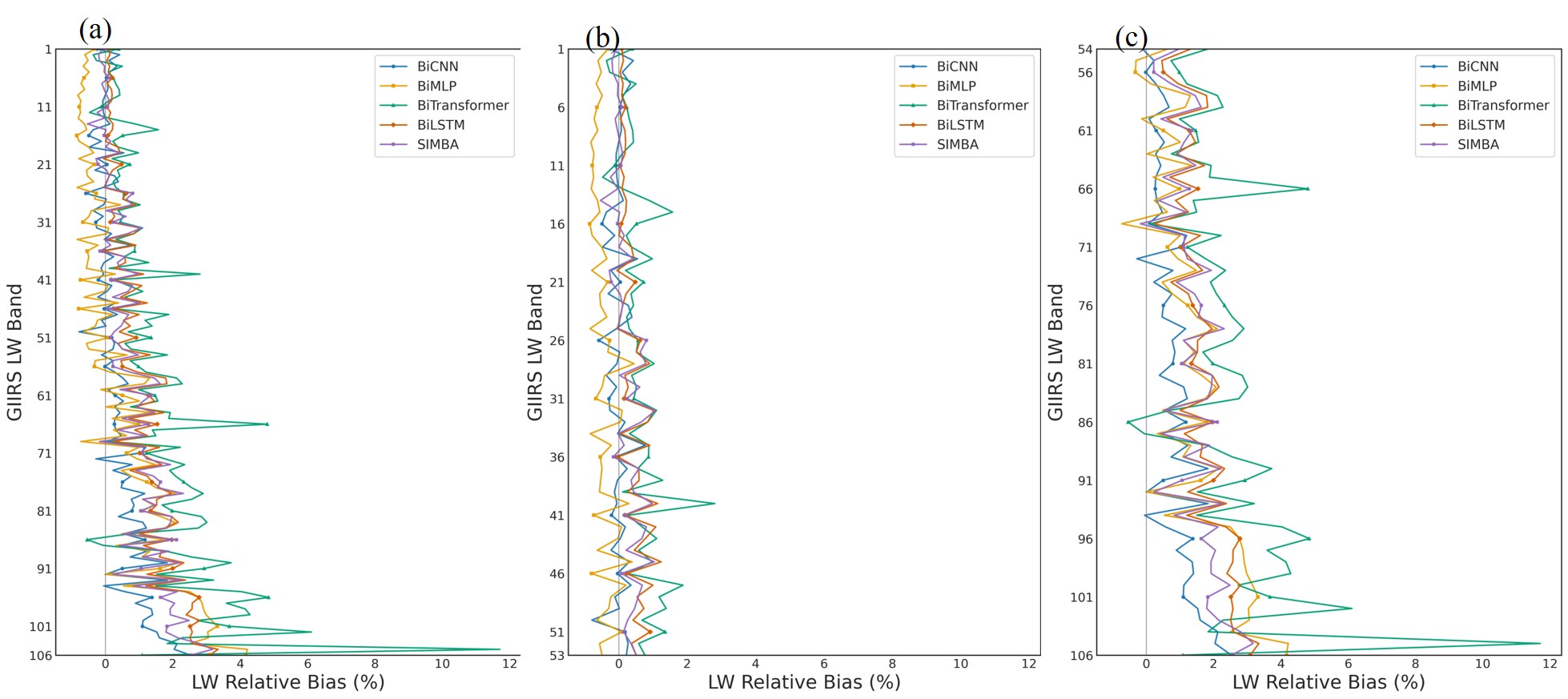}
\caption{Distribution of relative bias across 106 long-wave channels for different models: (\textbf{a}) full channel range; (\textbf{b}) enlarged view of CH1--CH53; (\textbf{c}) enlarged view of~CH54--CH106.}
\label{fig13}
\end{figure}

Figures~\ref{fig12} and~\ref{fig13} present the RMSE and relative bias distributions across the 106~LW channels, respectively. The~RMSE curves show that all models generally maintain relatively low errors in the window-region channels, whereas larger errors appear in channels associated with stronger absorption. SIMBA maintains lower or comparable RMSE values across most channels, with~a more gradual increase in error and smaller fluctuations, especially in channels that are more sensitive to middle- and upper-level atmospheric structures. The~relative bias curves further show that some baseline models exhibit larger oscillations or localized peaks in certain absorption-sensitive channels, whereas SIMBA remains generally more stable without obvious local~amplification.

\unskip

Overall, the~LW radiance reconstruction results demonstrate that SIMBA provides better observation-space consistency and more stable channel-wise performance than the comparison models, indicating a more reliable radiance--profile mapping for NWP-oriented~applications.

\subsubsection{Radiation of Medium-Wave~Bands}

From the perspective of NWP data assimilation, forward consistency in observation space is important for assessing whether the retrieved atmospheric states remain compatible with the observed radiance field in different spectral regimes. Therefore, the~MW radiance reconstruction results are analyzed in this~section.

\begin{figure}[H]

\includegraphics[width=1\textwidth]{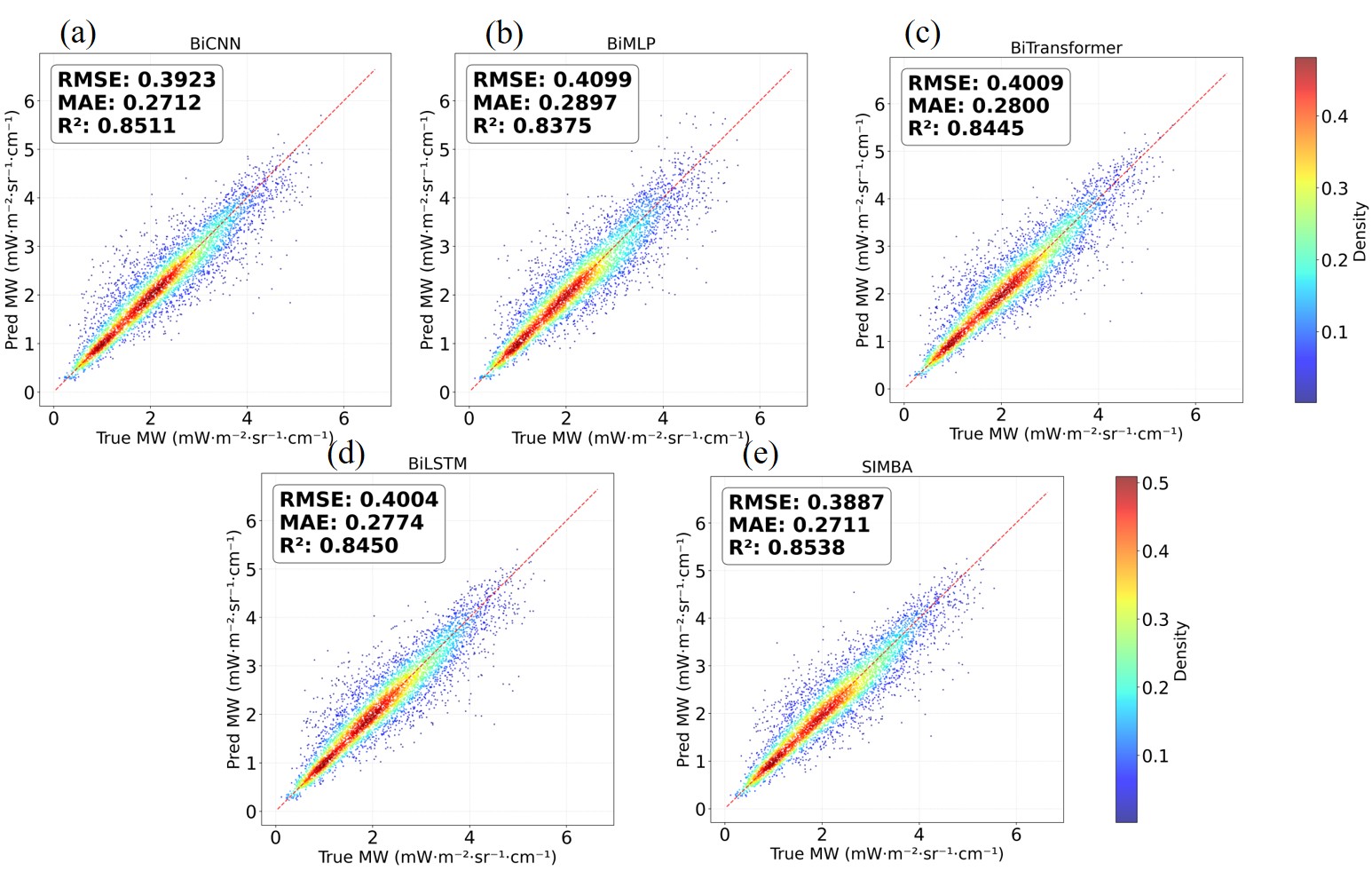}
\caption{Overall scatter plot comparison of medium-wave radiance reconstruction on the test set for different models: (\textbf{a}) BiCNN; (\textbf{b}) BiMLP; (\textbf{c}) BiTransformer; (\textbf{d}) BiLSTM; (\textbf{e}) SIMBA.}
\label{fig14}
\end{figure}

Figure~\ref{fig14} presents the overall scatter distributions of reconstructed MW radiances on the test set for different models. Overall, the~reconstructed values of all models follow the 1:1 reference line, indicating that the retrieved atmospheric profiles can reproduce the main characteristics of the observed MW radiances. Compared with the LW band, the~scatter distribution is generally more concentrated because of the smaller dynamic range of MW radiances, although~some dispersion remains in the higher-value region. SIMBA achieves the best overall performance in terms of RMSE and $R^2$, indicating slightly better consistency between the retrieved profiles and the observed radiances. Although~the numerical advantage is relatively moderate, the~more compact scatter pattern suggests that the proposed framework remains effective in preserving radiative consistency in the MW~band.

\begin{figure}[H]

\includegraphics[width=1\textwidth]{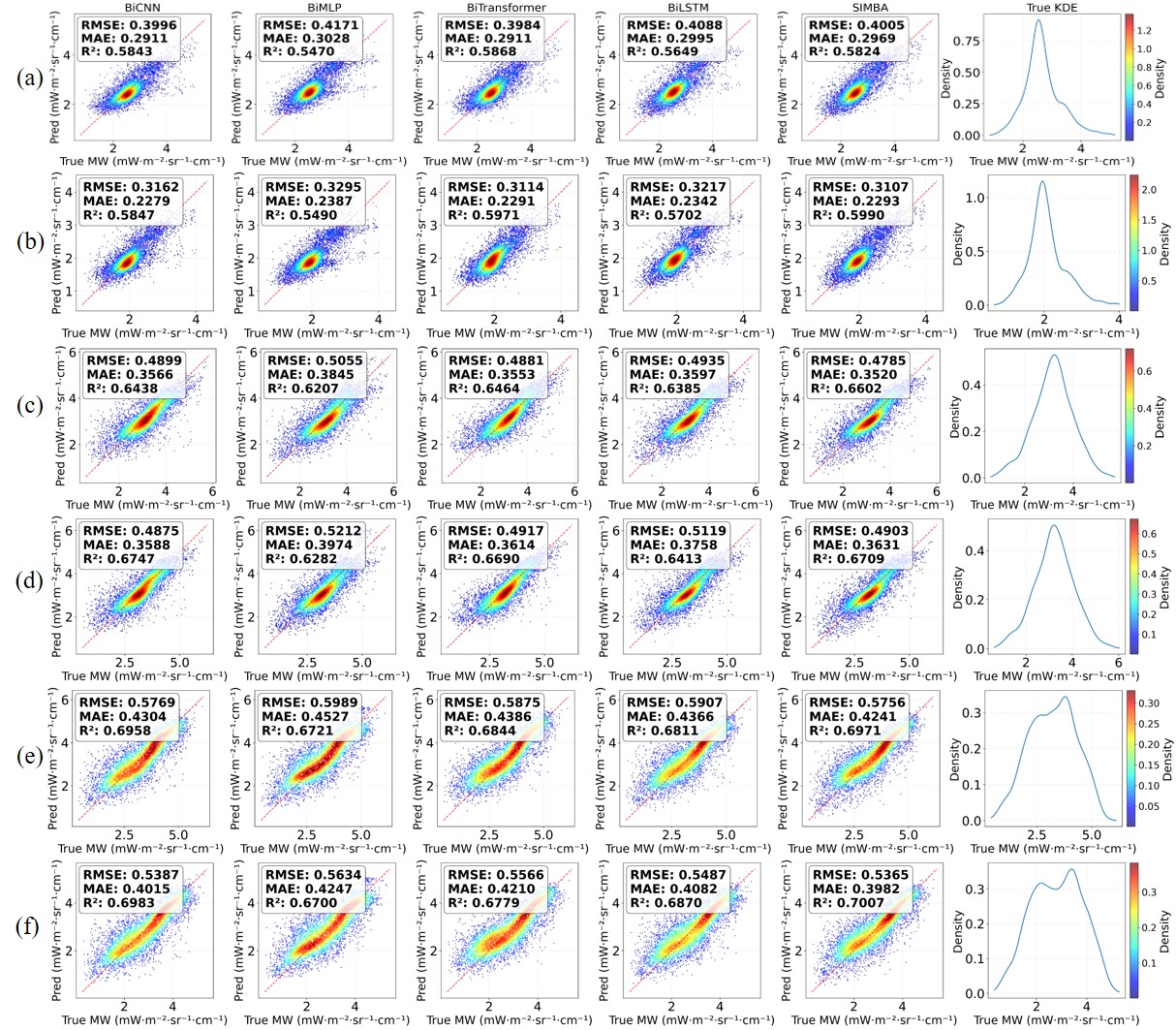}
\caption{Scatter plot comparison of medium-wave radiance reconstruction for six representative channels: (\textbf{a}) CH2; (\textbf{b}) CH20; (\textbf{c}) CH37; (\textbf{d}) CH42; (\textbf{e}) CH70; (\textbf{f}) CH88. Columns correspond to BiCNN, BiMLP, BiTransformer, BiLSTM, and~SIMBA, while the rightmost column shows the kernel density estimate (True KDE) of the observed radiance at each~channel.}
\label{fig15}
\end{figure}

To further analyze reconstruction differences across spectral positions, six representative channels (CH2, CH20, CH37, CH42, CH70, and~CH88) are selected from the 99 MW channels for detailed analysis in Figure~\ref{fig15}. In~the window-region channels (e.g., CH2 and CH20), all models exhibit relatively good linear relationships. In~the weak absorption-band channels (e.g., CH37 and CH42), the~scatter becomes more dispersed, reflecting the increased uncertainty associated with channels sensitive to the upper troposphere. In~relatively higher-level sensitive channels (e.g., CH70 and CH88), the~dispersion further increases for all models. Nevertheless, SIMBA maintains a narrower scatter range and a more gradual error growth trend across most representative channels, indicating stronger reconstruction stability in spectrally sensitive MW~regions.

\begin{figure}[H]

\includegraphics[width=1\textwidth]{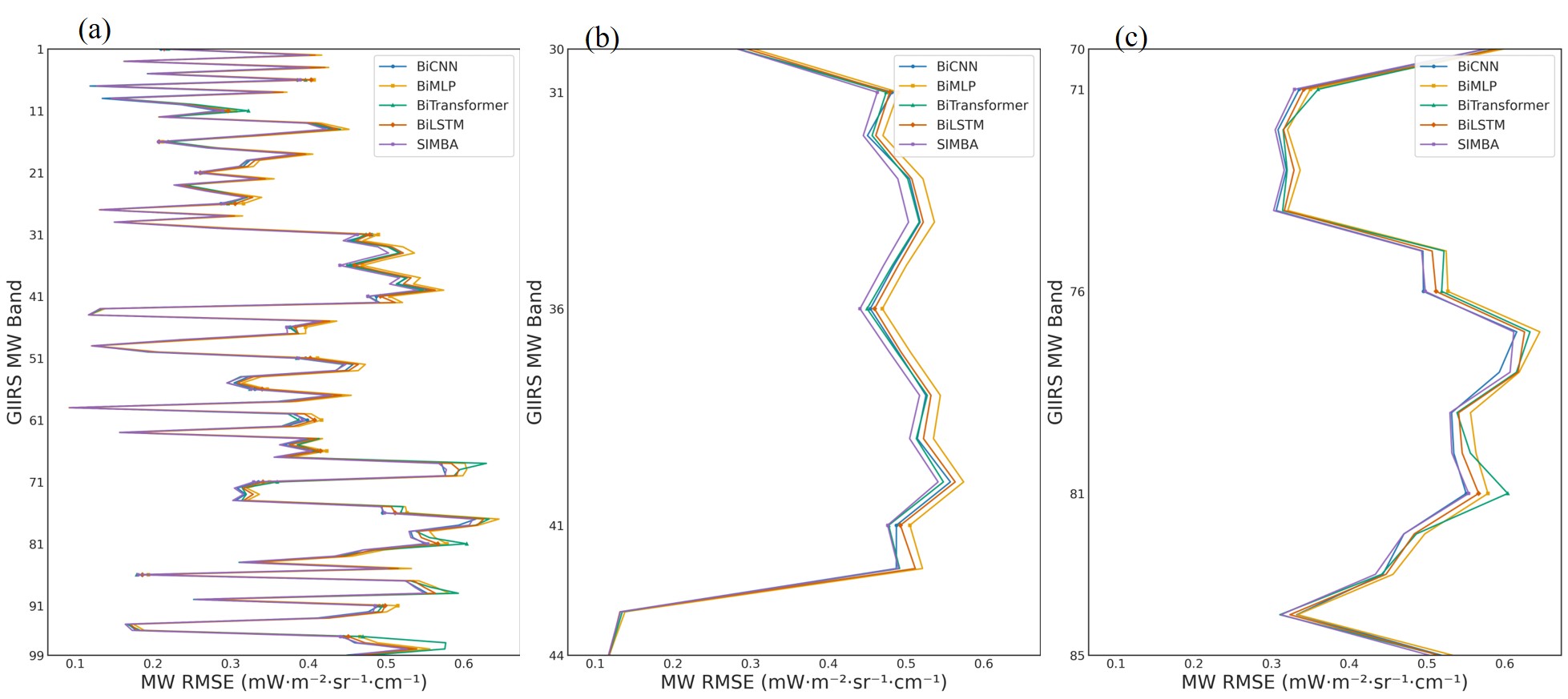}
\caption{Distribution of RMSE across 99 medium-wave channels for different models: (\textbf{a}) full channel range; (\textbf{b}) enlarged view of CH30--CH44; (\textbf{c}) enlarged view of~CH70--CH85.}
\label{fig16}
\end{figure}

\begin{figure}[H]

\includegraphics[width=1\textwidth]{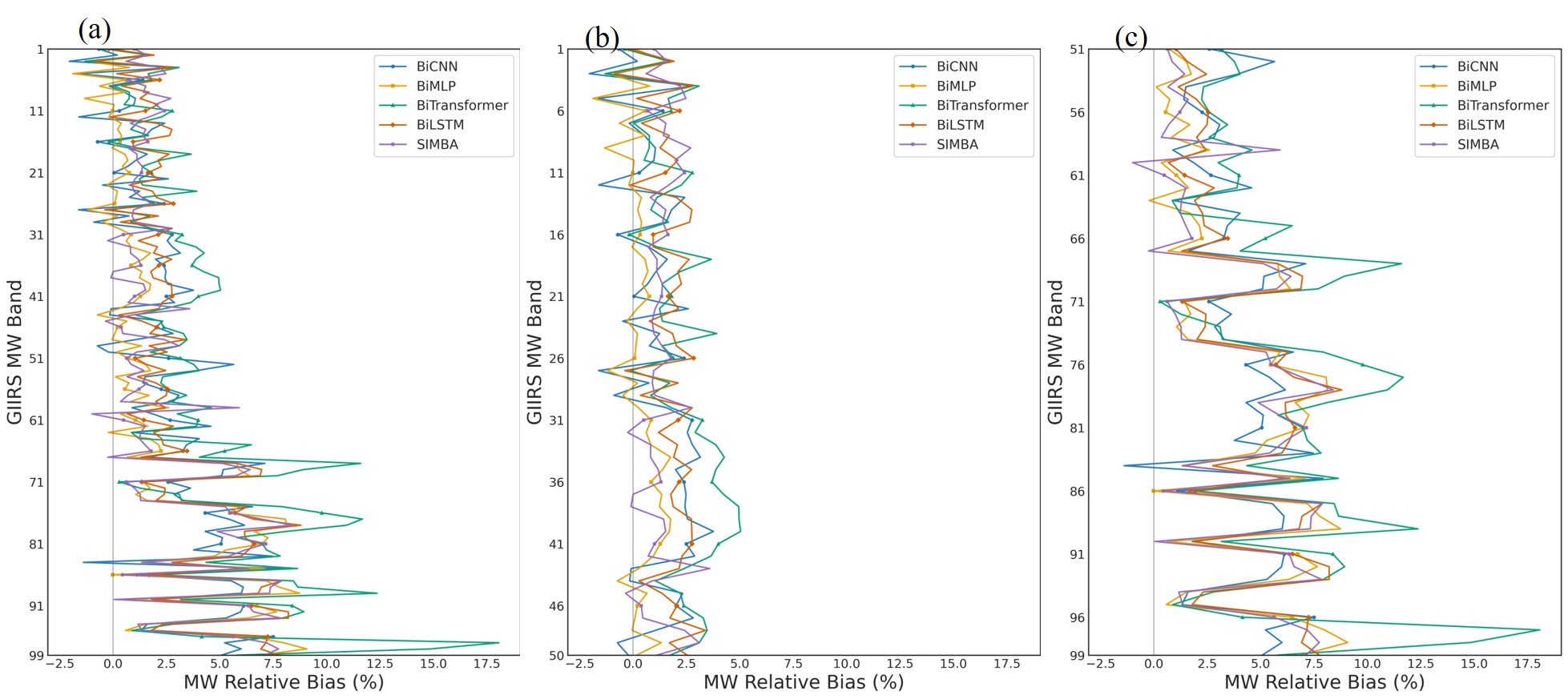}
\caption{Distribution of relative bias across 99 medium-wave channels for different models: (\textbf{a}) full channel range; (\textbf{b}) enlarged view of CH1--CH50; (\textbf{c}) enlarged view of~CH51--CH99.}
\label{fig17}
\end{figure}

Figures~\ref{fig16} and~\ref{fig17} present the RMSE and relative bias distributions across the 99 MW channels, respectively. Overall, all models exhibit relatively low errors in the window-region channels, whereas larger errors appear in weak absorption bands and in channels that are more sensitive to the middle and upper troposphere. The~channel-wise errors also show a certain oscillatory structure with increasing channel number. In~comparison, SIMBA maintains lower or comparable RMSE levels across most channels, with~relatively smaller fluctuations and without obvious localized amplification. The~relative bias curves further show that, although~fluctuations remain in certain channels, SIMBA is generally more stable than several baseline models and does not exhibit obvious local amplification in the medium- to high-numbered~channels.

Overall, the~MW radiance reconstruction results demonstrate that SIMBA provides better observation-space consistency and more stable channel-wise performance than the comparison models. This behavior is consistent with the trends observed in the LW band and further indicates a more reliable radiance--profile mapping for \mbox{NWP-oriented~applications}.

\vspace{-7pt}

\unskip

\subsection{Ablation~Study}

\textcolor{black}{To isolate the contribution of the bidirectional closed-loop design, ablation experiments compare three variants under the same Mamba backbone, data split, preprocessing, training protocol, and~evaluation metrics: Retrieval-Mamba (BT $\rightarrow$ profile, retrieval-only), Forward-Mamba (profile $\rightarrow$ BT, forward-only), and~SIMBA (bidirectional closed-loop with cycle consistency). Results are averaged over three repeated runs per setting.}

\textcolor{black}{As shown in Table~\ref{tab3}, SIMBA outperforms both baselines across all four variables. For~temperature retrieval, SIMBA achieves an MAE of 0.7141 (vs.\ 0.7372 for Retrieval-Mamba, roughly 3.1\% lower) and an RMSE of 1.0057 (vs.\ 1.0382, roughly 3.1\% lower). For~humidity, SIMBA reduces MAE from 0.2619 to 0.2527 (roughly 3.5\% lower) and RMSE from 0.5875 to 0.5677 (roughly 3.4\% lower) relative to Retrieval-Mamba. For~radiance reconstruction, SIMBA yields lower MAE and RMSE than Forward-Mamba on both LW (MAE: 3.9527 vs.\ 4.0335, roughly 2.0\% lower; RMSE: 6.0977 vs.\ 6.2830, roughly 2.9\% lower) and MW (MAE: 0.2712 vs.\ 0.2781, roughly 2.5\% lower; RMSE: 0.3879 vs.\ 0.3976, roughly 2.4\% lower), with~higher R$^2$ values in all cases. These results confirm that SIMBA's gains are not solely attributable to the Mamba backbone---coupling retrieval and forward simulation through the bidirectional closed-loop mechanism improves both retrieval accuracy and radiance reconstruction consistency.}

\begin{table}[H]

\setlength{\tabcolsep}{2.5pt}
\footnotesize
\caption{ \hl{Ablation} 
 results of SIMBA and baseline~models.}

\begin{adjustwidth}{-\extralength}{0cm}
\centering 
\begin{tabularx}{\fulllength}{lCCCCCCCCCCCC}
\toprule
\multirow{2}{*}{\vspace{-4pt}\textbf{Model}} & \multicolumn{3}{c}{\textbf{Temperature}} & \multicolumn{3}{c}{\textbf{Humidity}} & \multicolumn{3}{c}{\textbf{Long-Wave}} & \multicolumn{3}{c}{\textbf{Medium-Wave}} \\
\cmidrule{2-13}
& \textbf{MAE} & \textbf{RMSE} & \textbf{R}\boldmath{$^2$} & \textbf{MAE} & \textbf{RMSE} & \textbf{R}\boldmath{$^2$} & \textbf{MAE} & \textbf{RMSE} & \textbf{R}\boldmath{$^2$} & \textbf{MAE} & \textbf{RMSE} & \textbf{R}\boldmath{$^2$} \\
\midrule
Retrieval-Mamba & 0.7372 & 1.0382 & 0.9990 & 0.2619 & 0.5875 & 0.9888 & -- & -- & -- & -- & -- & -- \\
Forward-Mamba   & -- & -- & -- & -- & -- & -- & 4.0335 & 6.2830 & 0.8912 & 0.2781 & 0.3976 & 0.8470 \\
SIMBA           & \hl{\textbf{0.7141}} 
 & \textbf{1.0057} & \textbf{0.9991} & \textbf{0.2527} & \textbf{0.5677} & \textbf{0.9896} & \textbf{3.9527} & \textbf{6.0977} & \textbf{0.8975} & \textbf{0.2712} & \textbf{0.3879} & \textbf{0.8544} \\
\bottomrule
\end{tabularx}
\end{adjustwidth}
\label{tab3}
\end{table}
\unskip

\subsection{Discussion: Differentiable Forward Modeling and Remaining Validation~Requirements}

The above results show that SIMBA improves both atmospheric profile retrieval and radiance reconstruction through explicit coupling between atmospheric state space and radiance observation space. \textcolor{black}{From an NWP-oriented perspective, the~forward simulation branch can be regarded as a data-driven differentiable forward operator with RTM-like functionality under the evaluated conditions, rather than as a complete replacement for a physical radiative transfer model, as~illustrated in Figure S3.} It learns the mapping from atmospheric thermodynamic states to top-of-atmosphere radiances within a unified bidirectional framework:
\begin{equation}
\hat{\mathbf{P}}_i = F_{\theta}(\mathbf{R}_i,\mathbf{a}_i), \qquad
\hat{\mathbf{R}}_i = G_{\phi}(\mathbf{P}_i,\mathbf{a}_i).
\end{equation}

Because the forward branch is fully differentiable, its local sensitivity to atmospheric profile perturbations can be characterized by the Jacobian
\begin{equation}
\mathbf{J}_{G}(\mathbf{P}_i)=
\frac{\partial G_{\phi}(\mathbf{P}_i,\mathbf{a}_i)}{\partial \mathbf{P}_i}
=
\left[
\frac{\partial \hat{\mathbf{R}}_i}{\partial \mathbf{T}_i},
\frac{\partial \hat{\mathbf{R}}_i}{\partial \mathbf{q}_i}
\right],
\end{equation}
and the corresponding radiance response to a small perturbation $\delta \mathbf{P}_i$ can be approximated as
\begin{equation}
G_{\phi}(\mathbf{P}_i+\delta\mathbf{P}_i,\mathbf{a}_i)
\approx
G_{\phi}(\mathbf{P}_i,\mathbf{a}_i)+\mathbf{J}_{G}(\mathbf{P}_i)\,\delta\mathbf{P}_i.
\end{equation}
\hl{This} 
 formulation provides a possible basis for future Jacobian-related interpretation and channel-sensitivity analysis. However, differentiability alone does not demonstrate equivalence to a physical RTM or guarantee physically consistent sensitivity~estimates.

In this study, we focus on validating the bidirectional radiance–profile modeling capability of the proposed framework and its consistency in observation space. Nevertheless, rigorous validation using independent in situ observations and quantitative Jacobian comparisons with physical RTMs, such as RTTOV, is still required before the framework can be considered for operational assimilation-oriented~applications.

 \section{Conclusions}
 \label{sec4}
This study proposed SIMBA, a~simulation-constrained bidirectional Mamba framework for retrieving atmospheric temperature and specific humidity profiles from FY-4A/GIIRS observations. By~coupling retrieval (radiance $\rightarrow$ profile) and forward simulation (profile $\rightarrow$ radiance) within a unified architecture, SIMBA enables joint optimization across atmospheric state space and radiance observation space via a cycle-consistency mechanism, while the Mamba backbone captures long-range vertical dependencies in thermodynamic~profiles.

We evaluated SIMBA against several baseline models under both cloudy-sky and clear-sky conditions. Experimental results demonstrate that SIMBA consistently achieves the best RMSE, MAE, and~R$^2$ across temperature and specific humidity retrieval, and~yields the best LW and MW radiance reconstruction. Further analyses of vertical error distributions, channel-wise errors, and~ablation experiments confirm that the bidirectional closed-loop design improves both retrieval accuracy and radiance reconstruction stability. Overall, jointly constraining state space and observation space benefits hyperspectral infrared retrieval under complex radiative transfer conditions, and~provides a useful basis for NWP-oriented radiance--profile~modeling.

\textcolor{black}{Several limitations should be acknowledged. First, SIMBA should be regarded as a data-driven differentiable forward-modeling framework rather than a complete replacement for a physical RTM—residual retrieval and reconstruction errors remain non-negligible. Second, training and evaluation rely on ERA5 reanalysis as supervisory references. The~reported accuracy therefore reflects consistency with ERA5 rather than independent validation against radiosonde observations, and~the spatial-scale mismatch between GIIRS footprint and ERA5 grid may further introduce representativeness errors. Third, clear-sky and cloudy-sky conditions are not modeled separately. Both temperature and specific humidity retrieval accuracy under clear-sky conditions is lower than under cloudy skies (Table~\ref{tab2}b), indicating that separate modeling per sky condition could yield improvements.; evaluation across broader seasons and regions is also needed. Fourth, the~model achieved good fitting performance, but~the results lack physical interpretability. To~assess what the model has actually learned, we conducted a perturbation sensitivity analysis following the RTM approach (Figure S3). The~resulting channel sensitivities contain substantial high-frequency noise, which limits their physical interpretability and remains an open problem for future work.}

\textcolor{black}{Future work will address these limitations along several directions. Independent validation using radiosonde and other in situ observations will be conducted to assess model performance beyond ERA5 consistency. Separate modeling for clear-sky and cloudy-sky conditions will be explored, together with evaluation across broader seasons and regions. Longer time series will be incorporated to better capture seasonal and interannual variability. Finally, quantitative SIMBA--RTTOV Jacobian comparisons and noise-suppression techniques will be pursued to improve the physical interpretability of the learned \mbox{forward sensitivities}.}

\vspace{6pt}
\supplementary{\hl{~} 
}

\authorcontributions{Conceptualization, J.S. and F.W.; methodology, J.S.; software, J.S.; validation, J.S., H.H. and C.Y.; formal analysis, J.S.; investigation, J.S.; resources, X.L. and F.W.; data curation, J.S.; writing---original draft preparation, J.S.; writing---review and editing, J.S., H.H., C.Y., X.L. and F.W.; visualization, J.S.; supervision, X.L. and F.W.; project administration, F.W.; funding acquisition, X.L. and F.W. All authors have read and agreed to the published version of the manuscript.}

\funding{\hl{This} 
 research was funded by the National Natural Science Foundation of China    (Grant No. 42471437).}

\institutionalreview{Not applicable.}

\informedconsent{Not applicable.}

\dataavailability{The data presented in this study are available on request from the corresponding author. The~data are not publicly available due to privacy restrictions.} 

\acknowledgments{ We are grateful to the anonymous reviewers for their precious opinions and suggestions, whose professional reviews conspicuously enhanced the quality of this paper. Furthermore, our appreciation goes to the editorial team for their painstaking efforts and guidance throughout the
publication~process.}

\conflictsofinterest{The authors declare that they have no known competing financial interests or personal relationships that could have appeared to influence the work reported in this~paper.}


\begin{adjustwidth}{-\extralength}{0cm}
\reftitle{References}


\PublishersNote{}
\end{adjustwidth}

\end{document}